\documentclass[11pt]{article}
 \usepackage{amsmath,graphicx,amsfonts,amssymb,graphicx,epstopdf}

\newtheorem{theorem}{Theorem}[section]
\newtheorem{corollary}[theorem]{Corollary}
\newtheorem{lemma}[theorem]{Lemma}

\newcommand{\be}{\begin{equation}}
\newcommand{\ee}{\end{equation}}
\newcommand{\beq}{\begin{eqnarray}}
\newcommand{\eeq}{\end{eqnarray}}
\newcommand{\beqst}{\begin{eqnarray*}}
\newcommand{\eeqst}{\end{eqnarray*}}

\title{Finite lifespan of   solutions of the semilinear wave equation  in the  Einstein-de~Sitter  spacetime}

\author{Anahit Galstian and  Karen Yagdjian}

\begin{document}
\date{}
\maketitle

 \centerline{School of Mathematical and Statistical Sciences,}
   \centerline{University of Texas RGV,   Edinburg, TX 78539, U.S.A.}

\bigskip

\begin{abstract}
We examine the  solutions of the semilinear wave equation, and, in particular, of the $\varphi ^q$ model of quantum field theory in the curved space-time. 
More exactly, for $1< q<4$ we prove that the solution of the massless  self-interacting scalar field  equation in the  Einstein-de~Sitter universe
has finite lifespan.

\medskip

\noindent {\it Keywords:} Generalized Tricomi Equation;   Einstein-de~Sitter  spacetime;  Blowup of solution

\medskip

\noindent {Mathematics Subject Classification 2010:} 35C15; 35Q75; 35Q05; 83F05; 76H05

\end{abstract}

\section{Introduction}
\setcounter{equation}{0}
\renewcommand{\theequation}{\thesection.\arabic{equation}}

The equation for a self-interacting massless scalar field in the quantum field theory is the semilinear covariant wave equation
\begin{equation}
\label{IVPBOX}
\square_g \psi =  \lambda |\psi  |^{p-1}\psi \,,
\end{equation}
where $\square_g$ is a 
covariant d'Alembert's operator (the Laplace-Beltrami operator) in the  spacetime with the metric tensor $g$. The exponent $p>1 $ and the self-coupling constant $\lambda  $ show the intensity of    self-interaction.  
The metric of the Einstein~\&~de~Sitter universe (EdeS universe, see, e.g., \cite[p.123]{Choquet-Bruhat_book}, \cite[Sec.~5.3]{Hawking}) is a particular member of the
Friedmann-Robertson-Walker metrics
\begin{eqnarray*} 
ds^2= -dt^2+ a_{sc}^2(t)\left[ \frac{dr^2}{1-Kr^2}   + r^2  d\Omega ^2 \right]\,,
\end{eqnarray*}
where $K=-1,0$, or $+1$, for a hyperbolic, flat, or spherical spatial geometry, respectively.
For the EdeS universe the scale factor is $a_{sc}(t)=t^{2/3} $.
The covariant d'Alambert's operator,
\[
\square_g \psi = \frac{1}{\sqrt{|g|}}\frac{\partial }{\partial x^i}\left( \sqrt{|g|} g^{ik} \frac{\partial \psi }{\partial x^k} \right)\,,
\]
in the  EdeS spacetime is 
\begin{eqnarray*}
\square_{EdeS} \psi
& = &
-  \partial_t  ^2  \psi
-  2  t^{-1}\partial_t \psi  
+  t^{-\frac{4}{3}} A(x,\partial_x)\psi \,,
\end{eqnarray*}
where $ A(x,\partial_x)$ is a second order partial differential operator.

Thus, the equation for the self-interacting massless field in  the  Einstein-de~Sitter  spacetime  is the semilinear covariant wave equation
(\ref{IVPBOX}) 
which has singular  coefficients at $t=0$. The
covariant d'Alembert's operator  in the Einstein-de~Sitter spacetime belongs to
the family of the non-Fuchsian partial differential operators. The initial value problem for the equation (\ref{IVPBOX}) with the Cauchy data on hyperplane $t=0$ must be defined properly.   In \cite{Galstian-Kinoshita-Yagdjian} Galstian, Kinoshita and Yagdjian suggested such setting for the wave propagating in the  EdeS   spacetime when $A(x,\partial_x) $ is the Laplace operator on ${\mathbb  R}^n $.  In \cite{Galstian-Kinoshita-Yagdjian} the authors  introduced the weighted initial
value problem for the covariant (if $n=3$) wave equation and gave  explicit representation formulas for
the solutions. We generalize that setting and set   the problem for the semilinear equation  as follows  
\begin{eqnarray}
\label{WEEDS_A}
\begin{cases}
 \psi_{tt} - t^{-4/3}A(x,D_x)  \psi +   2   t^{-1}          \psi_t = F(\psi ),  
\qquad t>0 ,\,\, x \in {\mathbb R}^n,\cr
 \displaystyle   \lim_{t\rightarrow 0^+}\, t \psi  (x,t) = \varphi_0 (x), \quad x \in {\mathbb  R}^n , \cr
\displaystyle
\lim_{t\rightarrow 0^+}
  \left(  t \psi_t  (x,t) + \psi  (x,t)+3 t^{ - {1}/{3}} A(x,D_x) \varphi_0   (x   )  \right)
=  \varphi_1 (x),   \,\, x \in {\mathbb  R}^n \,,
\end{cases}
\end{eqnarray}
where $A(x,D_x) $ is an elliptic partial differential operator
$
A(x,D_x)= \sum_{|\alpha | \leq 2} a_\alpha (x) \partial_x^\alpha   
$  
with smooth real-valued coefficients $a_\alpha (x) \in C^\infty({\mathbb  R}^n )$, which are constant outside of some compact.
The two limits  of (\ref{WEEDS_A})  are taken in the sense of $H^1({\mathbb  R}^n) $ and $L^2({\mathbb  R}^n) $, respectively.

We define $p_{0}(n) $ as a positive root of the equation
\begin{eqnarray}
\label{pcr_intr}
   (n+3)p^2   -(n+13)p   -2 =0 
\end{eqnarray}
and  denote
 \begin{eqnarray*}
p_{cr}(n) := \max \left \{  p_{0}(n),1+\frac{6}{n} \right\}.
\end{eqnarray*}
Consider the operator 
\begin{eqnarray}
\label{1.5}
&  &
\label{3.12b_intr}
A(x,D_x)u = \frac{1}{a(x)}   \sum_{k,j=1,\ldots,n} \frac{\partial}{\partial x_k  } \left(  a_{k j} (x) \frac{\partial}{\partial x_j  } u \right)   \,,  
\end{eqnarray}
where $a (x), a_{k j} (x)   \in C^\infty ({\mathbb R}^n)$ and  
\begin{eqnarray}
\label{1.6}
&  &
\label{3.13_intr}
 a(x) \geq a_0 >0 ,\quad  a_{k j}(x)   =a_{jk }(x) \quad \mbox{\rm for all} \quad x \in  {\mathbb R}^n \,, k,j=1,2,\ldots,n\,,
\end{eqnarray}
with some  number $a_0$. Assume that the coefficients $a (x)$, $a_{k j} (x) $   are constant outside of some ball $B_{R_A}(0)$:
\begin{eqnarray}
\label{3.14_intr}
&  &
a_{k j}(x)   =c \delta _{jk }, \quad a(x)=1 \quad \mbox{\rm for all} \quad x \in  {\mathbb R}^n \,, |x| \geq R_A>0, \quad c>0,
\end{eqnarray} 
where $ \delta _{jk } $ is the Kronecker delta. 

We say that the solution $\psi \in C^2((0,T];{\mathcal D}'({\mathbb R}^n)) $ of the problem (\ref{WEEDS_A})
 obeys the finite propagation speed property if for every  point $(x_0,t_0) $ with $t_0>0$  and an open ball $B_R(x_0)=\{x \in {\mathbb R}^n\,;\, |x-x_0| <R\} $, the property 
\[
 \varphi_0 (x) =\varphi_1(x) =0  \quad \mbox{on}\quad   B_{R+ 3t_0^{1/3}s_A}(x_0)  \, , 
\]
implies
\[
\psi (x, t_0 ) =0  \quad \mbox{on}\quad  B_R(x_0) \,.
 \]
 Here
 \[
 s_A = \max_{ x \in {\mathbb R}^n, \, \xi  \in {\mathbb R}^n ,\,  |\xi|=1}  \frac{1}{a(x)}  \sum _{|\alpha | = 2} a_\alpha (x)\xi ^\alpha  \,.
 \] 

Although in quantum field theory the nonlinear term typically has a gauge invariant form $F(\psi )= |\psi|^{p-1} \psi  $, we will focus on  semilinear equations, which  
are  commonly used  models for general nonlinear problems (see \cite{YordanovSIAM,YagTricomi_GE} and the bibliography therein). Our first main result is the  following theorem. 
\begin{theorem} 
\label{T3.2}
Consider the problem (\ref{WEEDS_A}) with $F(\psi )=|\psi |^p$ and 
$A(x,D_x) $ being an elliptic operator  with the properties (\ref{3.12b_intr}),(\ref{3.13_intr}),(\ref{3.14_intr}). 
 If $ p>1$ and
\[
p<p_{cr}(n)\,,
\]
then for every arbitrary small number $\varepsilon >0 $ and an arbitrary number $s$ there exist functions  $\varphi_0 , \varphi_1  \in C_0^\infty({\mathbb  R}^n) $
with norms satisfying inequality 
\[
\|\varphi_0\|_{H_{(s)}({\mathbb  R}^n)}+ \|\varphi_1\|_{H_{(s)}({\mathbb  R}^n)} <\varepsilon
\]
such that the solution $\psi \in C((0,T);H^1({\mathbb  R}^n))\cap C((0,T);L^2({\mathbb  R}^n)) $ of the problem
(\ref{WEEDS_A}), which   obeys the finite propagation speed property, blows up in finite time. More precisely, there is $T<\infty$ such that 
\[
\lim_{t \nearrow  T } \int_{{\mathbb  R}^n} a(x) \psi (x,t)\,dx =\infty\,.
\]  
\end{theorem}

Note, for $n=3 $ we have $p_{cr}(3)=3$  that is the exponent of the $\varphi ^4$ model of quantum field theory.
The next corollary indicates that the equations (\ref{WEEDS_A}) possesses  global in time sign preserving solution only if   $p\geq 3$.   
\begin{corollary}
Assume that  $F(\psi  ) = | \psi |^{p-1}\psi  $, $1<p<3 $. For every arbitrary small number $\varepsilon >0 $ and an arbitrary number $s$ there exist functions  $\varphi_0 , \varphi_1  \in C_0^\infty({\mathbb  R}^n) $, 
$
\|\varphi_0\|_{H_{(s)}({\mathbb  R}^n)}+ \|\varphi_1\|_{H_{(s)}({\mathbb  R}^n)} <\varepsilon$,   
such that the positive solution $\psi \in C((0,T);H^1({\mathbb  R}^n))\cap C((0,T);L^2({\mathbb  R}^n)) $ of the problem (\ref{WEEDS_A})
 has a finite life-span.
\end{corollary} 
Note that, for the semilinear Klein-Gordon equation a global in time solvability is proved in 
\cite{Galstian-Yagdjian-NA} for the problem  with small initial data prescribed on the  hyper-surface $t=t_0> 0$ .   
\medskip

In Section~\ref{S5} we prove the finite propagation speed property for a subclass of operators of type (\ref{WEEDS_A}).
The next theorem shows that the blow up phenomenon is still present even if we remove the singularity at $t=0$  by shifting the initial hyperplane; the blow up   is caused by the semilinear term. 
 Consider the following Cauchy problem
\begin{eqnarray}
\label{6.16_intr}
\begin{cases}
 \psi_{tt} - t^{-2k}A(x,D_x)   \psi +   2   t^{-1}          \psi_t = |\psi |^p,  \qquad t>1 ,\,\, x \in {\mathbb R}^n,\cr
 \displaystyle
 \psi  (x,1) = \varphi_0 (x), \quad
  \psi_t  (x,1)
=  \varphi_1 (x),   \,\, x \in {\mathbb  R}^n ,
\end{cases}
\end{eqnarray}
where $k \in(0,1) $ and   $A(x,D_x) $ is an elliptic partial differential operator
with the properties (\ref{3.12b_intr}), (\ref{3.13_intr}), (\ref{3.14_intr}). 
Let $p_{0}(n,k) $ be a positive root of the equation  
\begin{eqnarray} 
\label{pcrII}
&  & 
p^2 (n+1 -k n)- p (2k+n+3-k  n )-2(1-k) = 0  \,.
\end{eqnarray}
The numbers $p_{0}(k) $ and  $p_{0}(n,k) $ can be regarded as an analog of the Strauss exponent that was defined for the semilinear wave equation in the Minkowski spacetime. (See, e.g., \cite{YordanovSIAM,YagTricomi_GE}.)

The equation of (\ref{6.16_intr}) is strictly hyperbolic for every bounded  interval of time and it has smooth coefficients. Consequently, for every smooth initial functions $\varphi_0  $ and $ \varphi_1 $ the problem (\ref{6.16_intr}) has the local solution.  According to the next theorem   a local in time solution, in general, cannot be prolonged to the global solution.    
\begin{theorem} 
\label{T6.9}
Assume that $ p>1$ and
\begin{equation}
\label{1.9}
 either \quad 1<p<1+\frac{2}{n (1-k)}\quad  or  \quad   1< p\leq     2+ \frac{2   k }{n+1-k n }  \,\, and \,\, p<p_{0}(n,k)\, .
\end{equation}
Then for every arbitrary small number $\varepsilon >0 $ and an arbitrary number $s$ there exist functions  $\varphi_0 , \varphi_1  \in C_0^\infty({\mathbb  R}^n) $
with norms satisfying inequality 
\begin{equation}
\label{1.10b}
\|\varphi_0\|_{H_{(s)}({\mathbb  R}^n)}+ \|\varphi_1\|_{H_{(s)}({\mathbb  R}^n)} <\varepsilon
\end{equation}
such that the solution $\psi \in C([1,T);H^1({\mathbb  R}^n))\cap C([1,T);L^2({\mathbb  R}^n)) $ of the problem
(\ref{6.16_intr}) that obeys the finite propagation speed property  blows up in finite time. More precisely, there is $T<\infty$ such that 
\[
\lim_{t \nearrow  T } \int_{{\mathbb  R}^n} a (x) \psi (x,t)\,dx =\infty\,.
\]  
\end{theorem}

Thus, according to this  theorem  for $n=3$  and $k= 2/3$   the blow-up occurs if $1<p<3 $. 
\begin{corollary}
Assume that  $F(\psi  ) = | \psi |^{p-1}\psi  $  and $p$ satisfies (\ref{1.9}). Then for every arbitrary small number $\varepsilon >0 $ and an arbitrary number $s$ there exist functions  $\varphi_0 , \varphi_1  \in C_0^\infty({\mathbb  R}^n) $, $ 
\|\varphi_0\|_{H_{(s)}({\mathbb  R}^n)}+ \|\varphi_1\|_{H_{(s)}({\mathbb  R}^n)} <\varepsilon$,  
such that the positive solution $\psi \in C([1,T);H^1({\mathbb  R}^n))\cap C([1,T);L^2({\mathbb  R}^n)) $ of the problem
(\ref{6.16_intr}) that obeys the finite propagation speed property  blows up in finite time. 
\end{corollary} 

In order to illustrate results of the theorems above we discuss below several examples which 
include, in particular, the Einstein-de~Sitter  spacetime of the {\it matter dominated universe}.
\smallskip

\noindent
{\bf Example 1.} Consider the covariant equation 
\begin{equation}
\label{1.10}
\psi_{tt} - t^{-4/3}\Delta  \psi +   2   t^{-1}          \psi_t = |\psi |^p,  
\qquad t>0 ,\,\, x \in {\mathbb R}^3,
\end{equation}
for the self-interacting waves propagating in the   Einstein-de~Sitter  spacetime.
Here $ \Delta $ is the Laplace operator in ${\mathbb  R}^3 $. 
According to Theorem~\ref{T3.2} and Theorem~\ref{T6.9} ($k=2/3$) if  $1<p<3 $, then for every arbitrary small number $\varepsilon >0 $ and an arbitrary number $s$ there exist functions  $\varphi_0 , \varphi_1  \in C_0^\infty({\mathbb  R}^3) $
with norms satisfying (\ref{1.10b}) 
such that the solution $\psi $ of the problem
(\ref{WEEDS_A}) or (\ref{6.16_intr}), respectively, for the equation (\ref{1.10}), which   obeys the finite propagation speed property, blows up in finite time. Note that $p =3$    is the exponent of the $\varphi ^4$ model of quantum field theory.
\medskip

The coefficients of the operator in the next examples depend   on the spatial variables as well. 
 
\noindent
{\bf Example 2.}  Consider the the  Einstein-de~Sitter  spacetime with the metric defined by 
\begin{eqnarray*} 
ds^2= -dt^2+ t^{4/3} \left[ \frac{dr^2}{1-Kr^2}   + r^2  d\Omega ^2 \right]\,,
\end{eqnarray*}
where $K=-1,0$, or $+1$. In the Cartesian coordinates $x=(x_1,x_2,x_3)$ this metric tensor is
\[
\left(
\begin{array}{cccc}
 -1 & 0 & 0 & 0 \\
 0 & t^{4/3}\frac{ 1- K \left(x_2^2+x_3^2\right)  }{1-K |x|^2)} &
  t^{4/3}\frac{   K x_1 x_2  }{1-K  |x|^2} 
&  t^{4/3}\frac{  K x_1 x_3 }{1-K  |x|^2} \\
 0 &  t^{4/3}\frac{  K x_1 x_2 }{1-K  |x|^2} 
& t^{4/3}\frac{ 1 -K \left(x_1^2+x_3^2\right) }{1-K |x|^2} 
&  t^{4/3}\frac{  K x_2 x_3 }{1-K  |x|^2} \\
 0 &  t^{4/3}\frac{ K x_1 x_3 }{1-K  |x|^2} 
&  t^{4/3}\frac{ K x_2 x_3 }{1-K  |x|^2} 
& t^{4/3}\frac{1  - K \left(x_1^2+x_2^2\right)}{1-K  |x|^2} \\
\end{array}
\right)
\]
and the  semilinear covariant wave equation in this metric reads
\begin{eqnarray}
\label{1.11b}
&  &
\psi _{tt}-t^{-4/3}{\mathcal A}(x, \partial_x ) \psi 
+ 2  t^{-1}\partial_t\psi  =  |\psi |^p\,,
\end{eqnarray} 
where
\begin{eqnarray}
\label{1.11c} 
{\mathcal A}(x, \partial_x )
& = &
( 1-Kx_1^2) \partial^2 _{x_1}  +(1-Kx_2^2)\partial_{x_2}^2   
+(1-Kx_3^2)\partial_{x_3}^2   -2 Kx_1 x_2 \partial_{x_1}\partial_{x_2} \\
&  &  
-2 Kx_1 x_3  \partial_{x_1}\partial_{x_3}   
-2K x_2 x_3 \partial_{x_2}\partial_{x_3}   
-3K x_1  \partial_{x_1} 
-3K x_2  \partial_{x_2}  
-3K x_3  \partial_{x_3} \,. \nonumber
\end{eqnarray} 
Thus, in the notation of (\ref{1.5}) we have 
\[
a_{kj}(x) = \frac{\delta_{kj}-K x_k x_j}{\sqrt{1-K |x|^2}}\,,\quad k,j=1,2,3\,,\qquad a(x)= \frac{1}{\sqrt{1-K |x|^2}}\,.
\]
We consider equation of (\ref{WEEDS_A}) that coincides with  (\ref{1.11b}) inside of the ball $B_R(0)  \subset {\mathbb R}^3$ and with 
(\ref{1.10}) outside of the ball $B_{2R}(0) $. The curvature ${\mathcal R}$ of such spacetime is 
\[
{\mathcal R}=\frac{4}{3 t^2}+\frac{6 K}{t^{4/3}} \quad \mbox{\rm in } \quad B_R(0)  \quad \mbox{\rm while} \quad  {\mathcal R}=\frac{4}{3 t^2}  \quad \mbox{\rm in } \quad   ({\mathbb R}^3\setminus B_{2R}(0))\,.
\]
In oder to make coefficients of this operator more explicit in $B_{2R}(0) \setminus B_R(0)$  one can use the standard cut-off function $\chi =\chi (x) \geq 0 $ and attach to $K$ the factor   
$\varepsilon  \chi (x)$. For sufficiently small $\varepsilon >0 $ the  conditions  (\ref{3.12b_intr}),(\ref{3.13_intr}),(\ref{3.14_intr}) are fulfilled. Another equation satisfying all conditions is the following one 
\begin{eqnarray*}
&  &
\psi _{tt}-t^{-4/3}\frac{1}{1+a(x)\chi (x)}{\mathcal A}(x, \partial_x ) \psi 
+ 2  t^{-1}\partial_t\psi  =  |\psi |^p\,,
\end{eqnarray*} 
where $ a(x)$ is any smooth non-negative function and the operator $ {\mathcal A}(x, \partial_x )$ is given by (\ref{1.11c}) inside of $B_R(0) $ and is $\Delta  $ outside of $B_{2R}(0)$.
Then all conclusions of  Example~1 are valid also for these equations. 
\smallskip

\noindent 
{\bf Example 3.} Consider now problem (\ref{6.16_intr}) with $A(x,D_x)=\Delta  $.
For   the   radiation dominated universe   $k=1/2$ and $n=3$. The first case   of  (\ref{1.9}) in Theorem~\ref{T6.9} reads $  1<p<  \frac{7}{3} \,.$  We obtain $ 1<p\leq 2 $ from the second one. 
Thus, for the equation (\ref{6.16_intr})  there is a blowing up  small data   solution if $  1<p<  \frac{7}{3} $. Another example  can be obtained by replacing $\Delta  $ with ${\mathcal A}(x,\partial_x) $ (\ref{1.11c}) inside of some ball in $ {\mathbb R}^3$ and with the modification  similar to the one has used in  
Example~2.
\medskip

Next two examples have spacetimes with non flat spatial slices.   

\noindent
{\bf Example 4.} Let a spacetime be defined by the following metric  
\[
ds^2= -dt^2+ t^{2k}\left( \frac{\beta }{x^2+1}dx^2 + \frac{x^2+1}{\beta } dy^2+ dz^2\right)
\]
inside of some ball, where $k $ and $\beta > 0 $ are real numbers. The curvature of this spacetime is $ 12 k^2 t^{-2}- 2 t^{-2k} \beta^{ -1}- 6 k t^{-2} $, while the spatial slices have the constant  curvature  $- 2 \beta^{-1 } $. We consider  semilinear equation in this spacetime  
\begin{eqnarray}
\label{1.11}  
 \psi_{tt} - t^{ -2k}\left(\frac{x^2+1}{\beta } \partial_x^2 \psi   +  \frac{2x }{\beta } \partial_x\psi   + \frac{\beta } {x^2+1}\partial_y^2\psi  + \partial_z^2 \psi\right)  \psi +   2 t^{-1}          \psi_t = |\psi |^p,  \quad t>0 .  
\end{eqnarray}
The modification outside of some ball is similar  to the one mentioned in Example~2. The  equation (\ref{1.11}) is a covariant wave equation if $k=2/3 $.  It is easy to verify that    Theorem~\ref{T6.9} can be applied to the problem for this equation.   
\smallskip

\noindent
{\bf Example 5.} Consider the spacetime with the metric  
\[
ds^2= -dt^2+ t^{2k}\left(\frac{\beta   }{e^{-x^2}+1}  dx^2 + \frac{e^{-x^2}+1}{\beta } dy^2+ dz^2\right)
\]
inside of some ball, where $k $ and $\beta >0 $ are real numbers. The curvature of the spacetime is  
\[
  2  \beta^{-1}  t^{-2 (k+1)} \left(3 k (2 k-1) \beta    t^{2 k}+t^2 e^{-x^2}\left(1-2 x^2\right)\right) ,
\] 
 while the spatial slices have the curvature  $  2 \beta^{-1}t^{-2 k} e^{-x^2} \left(1-2 x^2\right)  $. Theorem~\ref{T6.9}  can be applied to the semilinear equation of (\ref{6.16_intr}) in this spacetime,    
where 
\begin{eqnarray*}  
A(x,D )\psi  
& = &
 \frac{e^{-x^2}+1}{\beta } \partial_x^2 \psi   +  \frac{-2x e^{-x^2}}{\beta } \partial_x\psi   + \frac{\beta } {e^{-x^2}+1}\partial_y^2\psi  + \partial_z^2 \psi\,.
\end{eqnarray*}
The    equation (\ref{6.16_intr}) in this spacetime is a covariant wave equation if $k=2/3 $.
It will be interesting to replace requirement on the coefficients of $A(x,D ) $  to be constant outside of a ball with a condition on their rate of convergence to the constants    at  infinity. 
 
The last two examples belong to   more general class of equations  written in the background given by the following  metric 
\[
ds^2= -dt^2+t^{2k}\left(G_1(x,y,z) dx^2 + G_2(x,y,z) dy^2+ G_3(x,y,z)dz^2\right) 
\]
 such that $G_1(x,y,z)    G_2(x,y,z)   G_3(x,y,z)=constant \not=0$.
\medskip

This paper is organized as follows. In Section \ref{S2} we introduce the basic  ideas of the proof of  Theorem~\ref{T3.2} and give main tools which will be also used in the next sections.  In Section~\ref{S3} we prove Theorem~\ref{T6.9}. The existence of the local in time solution in proved in Section \ref{S4}. Section~\ref{S5} is devoted to the uniqueness problem and the finite speed of propagation property.

\section{Proof of Theorem~\ref{T3.2}}
\setcounter{equation}{0}
\renewcommand{\theequation}{\thesection.\arabic{equation}}

\label{S2}

The number  $p_{0}(n) $ is defined as a positive root of the equation (\ref{pcr_intr}), 
that is, 
\begin{eqnarray*}
p_{0}(n)= \frac{n+13 + \sqrt{n^2+34n+193}}{2(n+3)}\,.
 \end{eqnarray*}
 It is easily   seen that 
 \begin{eqnarray*}
 p_{0}(n) < \frac{2n+10}{n+3} \quad \mbox{\rm for all} \quad n \geq 4,
\end{eqnarray*}
and that 
 \begin{eqnarray*}
 p_{0}(n) < 1+\frac{6}{n} \quad \mbox{\rm for all} \quad n \leq  4.
\end{eqnarray*}
If we denote
\begin{equation}
\label{OPLS}
{\mathcal L} :=\partial_{t}^2
-  t^{-4/3} A(x,D_x)      + 2 t^{-1 }        \partial_t  ,\qquad
{\mathcal S}:= \partial_{t}^2
-  t^{-4/3} A(x,D_x)   \,,
\end{equation}
then we can easily check for $t \not= 0$ the following operator identity
\begin{equation}
\label{OPI}
 t^{-1} \circ {\mathcal S}  \circ t
 =
   {\mathcal L}   \,.
\end{equation}
The last equation suggests  a partial Liouville  transform of an unknown function $\psi $ with $u$  
\[
\psi =t^{-1}u  .
\]
 Then the problem for $u$ is:
\begin{eqnarray} 
\label{ucl}
\begin{cases}
 u_{tt} - t^{-4/3} A(x,D_x)    u  = t^{1-p} |u|^p,  \qquad t>0 ,\,\, x \in {\mathbb R}^n,\cr
 \displaystyle   \lim_{t\rightarrow 0^+}\, u (x,t) = \varphi_0 (x), \qquad  \,\, x \in {\mathbb R}^n, \cr
\displaystyle
\lim_{t\rightarrow 0^+}
  \left(  u _t (x,t)+3 t^{ - {1}/{3}} A(x,D_x)  \varphi_0  (x   )  \right)
=  \varphi_1 (x),   \quad x \in {\mathbb  R}^n \,.
\end{cases}
\end{eqnarray}
Recall (\ref{1.5}), (\ref{1.6}),   
and that the coefficients $a (x) , a_{k j} (x)\in C ^\infty ({\mathbb R}^n) $ are constant outside of some ball $B_{R_A}(0)$.
Denote
\[
F (t)= \int_{{\mathbb R}^{n}} a(x) u(x,t) \,dx \, .
\]
Then $F\in C^2(0,T) $ provided that the function $u$ is defined for all $(x,t) \in {\mathbb R}^n\times (0,T) $, and
\begin{eqnarray*} 
 \displaystyle   \lim_{t\rightarrow 0^+}\, F(t)= 
  \int_{{\mathbb R}^{n}}  a(x) \varphi_0 (x)  \,dx =C_0 ,
\end{eqnarray*}
while  
\begin{eqnarray*} 
\lim_{t\rightarrow 0^+} \, F'(t) 
& = & 
 \lim_{t\rightarrow 0^+}\,  \int_{{\mathbb R}^{n}} a(x) \left[ u _t (x,t)+3 t^{ - {1}/{3}} A(x,D_x) \varphi_0  (x   )
-3 t^{ - {1}/{3}} A(x,D_x) \varphi_0  (x   )\right]\,dx   \\
& =   &
 \lim_{t\rightarrow 0^+}\,  \int_{{\mathbb R}^{n}} a(x) \left[ u _t (x,t)+3 t^{ - {1}/{3}} A(x,D_x)  \varphi_0  (x   )  \right]\,dx = \int_{{\mathbb R}^{n}}  \,a(x)  \varphi_1 (x)\,dx =C_1\,.     
\end{eqnarray*}
Thus
\[
F \in C^1[0,\infty)\cap C^2(0,\infty)\,.
\]
From the  equation we have
\begin{eqnarray*}
F ''
& = &
t^{1-p} \int_{{\mathbb R}^{n}}a(x)  |u(x,t)|^p \,dx \geq 0 \quad \mbox{for all}\quad t>0.
\end{eqnarray*}
Furthermore,
\begin{eqnarray*}
F (t)
&= &
F (\varepsilon  )
+ (t-\varepsilon ) F' (\varepsilon  )  +\int_\varepsilon ^t   \int_\varepsilon ^{t_1} F'{}'(t_2 )d t_2  \, d t_1  \\
&\geq  &
F (\varepsilon  )
+ (t-\varepsilon ) F' (\varepsilon  )    \geq 0 \quad \mbox{for all}\quad t\geq \varepsilon \,.
\end{eqnarray*}
By letting $\varepsilon \to 0^+ $ we obtain
\begin{eqnarray*}
F (t)
& \geq  &
t C_1+C_0\geq 0 \quad \mbox{for all}\quad t\geq 0 
\end{eqnarray*}
provided that $C_0   \geq 0 $ and $C_1  \geq 0 $. We can assume also that supp~$\varphi _i \subseteq 
B_R(0):=\{x \in {\mathbb R}^n \,|\, |x| \leq R\}$, $i=0,1$ and $R\geq R_A $.  On the other hand,
using the compact support of $u(\cdot,t)$ and H\"older's inequality we obtain 
with  $\phi (t)=3t^{1/3} $
\begin{eqnarray*}
\left| \int_{{\mathbb R}^n} a(x) u (x,t) \,dx \right|^{p}
& \lesssim  &
(R+  \phi (t))^{ n(p-1)}
\left( \int_{|x| \le R+  \phi (t)} a(x) |u (x,t)|^{p} \,dx \right) \,.
\end{eqnarray*}
Here and henceforth, if $A$ and $B$ are two non-negative quantities, we use $A \lesssim  B$ ($A \gtrsim  B $)
to denote the statement that $A\leq CB $ ($AC\geq  B$) for some absolute constant $C>0$. Hence
\begin{eqnarray}
\label{fdtdt}
F '' (t)
& \gtrsim &
   (R+ \phi (t))^{ - (n+3)(p-1)}  |F(t)|^p\quad \mbox{for all}\quad t\geq 0 \,. 
\end{eqnarray}
If \, $1< p<1 +\frac{6}{n}$ \, and \,$C_1>0$,  then we can apply Kato's lemma (see, e.g., \cite[Lemma~2.1]{Yag_2005}) since
\[
p-1>\frac{(n+3)(p-1)}{3}-2\Longleftrightarrow p<\frac{6}{n}+1 
\]
that proves that solution blows up for such $p$.  
\smallskip

Next  we consider the case of  {$  1< p\leq  (2n+10)/(n+3)$ \,and \,$p<p_{0}(n) $}. For   $\varphi _0 \in C_0^{[\frac{n}{2}]+3} ({\mathbb R}^n) $, according to \cite{JDE2015}, the solution of the  problem
\begin{eqnarray}
\label{phy10}
 \begin{cases} \displaystyle  {\mathcal S}  u  = 0,  \qquad x \in {\mathbb R}^n,\quad t>0,\cr
 \displaystyle  \lim_{t\rightarrow 0}u (x,t) = \varphi_0 (x),  \qquad
 \displaystyle   \lim_{t\rightarrow 0}
\Big (
  u_t  (x,t) +3 t^{   - {1}/{3}  } A(x,D_x) \varphi_0
   (x   )
  \Big )
=0,  \qquad x \in {\mathbb R}^n \,,
\end{cases}
\end{eqnarray}
  is given by  the function
\begin{eqnarray*}
u (x,t)
& = &
  v_{\varphi_0}  (x, 3t^{1/3})  - 3t^{1/3}   ( \partial_r   v_{\varphi_0} )  (x, 3t^{1/3}) \,,
\end{eqnarray*}
 where $v_{\varphi  }   ( x,3t^{1/3} ) $ is the value of the solution $v   ( x,r  ) $ to the Cauchy problem 
\[
v_{rr} -    A(x,D_x) v =0  ,  \,\,
 v(x,0) = \varphi  (x) , \,   v_{t } (x,0) =  0,
\]
taken at the point $ ( x,r  )=( x,3t^{1/3})$. Hence, if we assume that $A(x,D_x) \varphi =\varphi $, 
 then we obtain
\[
v_\varphi (x,t)= \cosh (t)\varphi (x)
\]
and, consequently,
\[
u (x,t)
  =  
\left( \cosh (3t^{1/3})     - 3t^{1/3}  \sinh (3t^{1/3}) \right) \varphi (x) \,.
\]
The second independent solution with separated variables is
\[ 
w (x,t)
  =  
\left( \sinh (3t^{1/3})     - 3t^{1/3}  \cosh (3t^{1/3}) \right) \varphi (x) \,.
\]
Thus, the function $v(x,t)=u(x,t)
-  w(x,t)  $, that is,
\begin{eqnarray*}
v(x,t) =
 \left( 3 \sqrt[3]{t} +1  \right)\exp \left( -3 \sqrt[3]{t}    \right)\varphi (x)=  \left( \phi (t) +1  \right)\exp \left( -\phi (t)    \right)\varphi (x)
\end{eqnarray*}
solves the problem (\ref{phy10}) with $\varphi _0 = \varphi  $.  Moreover, $ v$ is such that
\begin{eqnarray*}
v(x,0)
  =
\varphi (x) \,,\qquad \lim_{t \to \infty} v(x,t) =0  \,.
\end{eqnarray*}
The following lemma generalizes corresponding result from  \cite{YordanovSIAM}.
\begin{lemma}
There is a smooth function $\varphi (x) $  such that 
\[
A(x,D_x) \varphi (x) = \varphi (x)  \quad  \mbox{for all} \quad   x \in {\mathbb R}^n
\]
and 
\[
\varphi (x) = \int_{{\mathbb S}^{n-1}}e^{x \omega  }\,d\omega \quad  \mbox{for all} \quad x, \quad   |x| \geq R_A+1\,.
\]
Moreover, 
\begin{eqnarray*} 
\varphi (x) \sim C_n|x|^{-(n-1)/2}e^{|x|}\quad \mbox{as} \quad |x| \to \infty\,.
\end{eqnarray*} 
\end{lemma}
\medskip

\noindent
{\bf Proof.}
We have 
\[
\Delta  \int_{{\mathbb S}^{n-1}}e^{x \omega  }\,d\omega= \int_{{\mathbb S}^{n-1}}e^{x \omega  }\,d\omega  \quad  \mbox{for all} \quad   x \in {\mathbb R}^n\,,
\]
where $\Delta  $ is the Laplace operator. It is well known \cite{YordanovSIAM} that  
\begin{eqnarray*} 
\varphi _L (x):= \int_{{\mathbb S}^{n-1}}e^{x \omega  }\,d\omega \sim C_n|x|^{-(n-1)/2}e^{|x|}\quad \mbox{as} \quad |x| \to \infty\,.
\end{eqnarray*}   
To find the function $\varphi (x) $ we solve the Dirichlet problem for the elliptic 
equation 
\[
A(x,D_x) \varphi (x) - \varphi (x) =0 \quad \mbox{\rm in} \quad B_{ R_A+1}(0) \,, \quad \varphi (x) =  \varphi _L (x)  
\quad \mbox{\rm on} \quad \partial B_{ R_A+1}(0)= {\mathbb S}^{n-1}_{ R_A+1 }
\] 
(see, e.g. \cite[Sec~9.6]{Gilbarg-Trudinger}). We set also $  \varphi (x)= \varphi_L (x)$
if $|x| \geq  R_A$.  The lemma is proved.  \hfill $\square$
\bigskip

Thus, the function $v(x,t)$ is the ``low frequency'' solution of the linear equation
\[
v_{tt}- t^{-4/3}A(x,D_x) v =0\,.
\]
Next we define the function $F_1(t)$,
\begin{eqnarray*}
 F_1(t)
& := &
\int_{{\mathbb R}^n} a(x) u (x,t) v(x,t) \,dx \,,
\end{eqnarray*}
that is, the projection of the solution on the ``low frequency''    eigenspace of the problem for the operator $A(x,D_x)$. Here $F_1\in C^2(0,T) $. We estimate the function $F_1 $  from above as follows
\begin{eqnarray}
\label{3.12}
F''(t)
& \gtrsim  &
\left( \int_{|x| \le R+  \phi (t)} |v(x,t)|^{p/(p-1)} \,dx \right)^{1- p }
t^{1-p}\left| F_1(t) \right|^{p}\,.
\end{eqnarray}
To find out  the properties of $F_1(t) $ we need the following lemma.
\begin{lemma} 
\label{L6.2}
The function
\begin{eqnarray*}
 \lambda (t)
& = & 
\left( \phi (t) +1  \right)\exp \left( -\phi (t)    \right)
  \end{eqnarray*}
  solves the equation
\[
\lambda ''(t)-t^{-\frac{4}{3}}\lambda (t)=0
\]
and has the following properties:
\begin{eqnarray*}
(i)\,\, \lambda' (t)
& = &
-\frac{9}{  \phi (t)} \exp \left( -\phi (t)    \right)\leq 0\,,\\
(ii)\,\,
\,\, \lim_{t \to 0}  \lambda (t)
& = &
1\,,\qquad
\lim_{t \to \infty}  \lambda (t)   = 0\,, \qquad
\lim_{t \to \infty}  \lambda' (t)
  =   0\,,\\
(iii)\,\,\frac{ \lambda' (t) }{ \lambda (t) }
& = &
-\frac{9 }{ \phi (t)\left( \phi (t) +1  \right) }\,.
\end{eqnarray*}
\end{lemma}
\medskip

\noindent
{\bf Proof.} It can be verified  by straightforward calculations. \hfill $\square$
\medskip

Next we turn to the function $\varphi (x) $. The following lemma is an analog of Lemma 2.3~\cite{YordanovSIAM}.

\begin{lemma} 
\label{L2.3}
Assume that $p>1$.
Then
\begin{eqnarray*} 
  \int_{|x| \le \tau }|\varphi  (x)| ^{p/(p-1)} \,dx
 \leq
c_R   \tau ^{ \frac{n-1}{2}\frac{p-2}{p-1} }e^{\tau  \frac{p}{p-1}} \quad \mbox{for all} \quad \tau \geq R_A+1. 
\end{eqnarray*}
\end{lemma}
\medskip

\noindent
{\bf Proof.} Indeed, for $\tau \geq R_A+1 $ we have  
\begin{eqnarray*} 
  \int_{|x| \le \tau }|\varphi  (x)| ^{p/(p-1)} \,dx  
& =  &
\int_{|x| \le R_A+1 }|\varphi  (x)| ^{p/(p-1)} \,dx +   \int_{ R_A+1 \leq |x| \le \tau }|\varphi  (x)| ^{p/(p-1)} \,dx \\
& =  &
\int_{|x| \le R_A+1 }|\varphi  (x)| ^{p/(p-1)} \,dx +   \int_{ R_A+1 \leq |x| \le \tau }|\varphi_L  (x)| ^{p/(p-1)} \,dx \\
& =  &
\int_{|x| \le R_A+1 }|\varphi  (x)| ^{p/(p-1)} \,dx +   \int_{ |x| \le \tau }|\varphi_L  (x)| ^{p/(p-1)} \,dx  \,.
\end{eqnarray*}
The application of Lemma 2.3~\cite{YordanovSIAM} completes the proof. 
\hfill $\square$

\begin{lemma}
Assume that $\, \varphi_0, \varphi_1 \in C_0^\infty({\mathbb R}^{n})$,    and that
\begin{eqnarray*}
 &  &
   \int_{{\mathbb R}^n}  a(x) \varphi_1 (x) \varphi (x)\,dx
 \geq 18    \int_{{\mathbb R}^n}  a(x)\varphi_0 (x)
 \varphi (x)\,dx  >0  \,,
\end{eqnarray*}
then
\begin{eqnarray*}
    F_1(t)
& \gtrsim   &
(9\sqrt[3]{t^2}-1)     \int_{{\mathbb R}^n}   a(x) \varphi_1 (x)\varphi (x)\,dx  \quad \mbox{for all} \quad t>1\,.
\end{eqnarray*}
\end{lemma}
\medskip

\noindent
{\bf Proof.} We have
\[
F_1(0) = \lim_{t \to 0^+} \int_{{\mathbb R}^n}  a(x)u (x,t) v(x,t)\,dx =   \int_{{\mathbb R}^n}  a(x)\varphi _0 (x) \varphi (x) \,dx \geq c_0 >0\,.
\]
For every $\epsilon >0 $ we have
\begin{eqnarray*}
0
& = &
\int_\varepsilon ^t \int_{{\mathbb R}^n}  a(x)( u_{tt} (x,\tau ) -\tau ^{ - {4}/{3}}A(x,D_x) u -\tau ^{1-p}|u|^p) v(x,\tau )\,dx \,d\tau  \\
& =  &
\int_\varepsilon ^t \int_{{\mathbb R}^n}  a(x)  u_{tt} (x,\tau )  v(x,\tau )\,dx \,d\tau    -\int_\varepsilon ^t \int_{{\mathbb R}^n}  \tau ^{ - {4}/{3}}u(x,\tau )  a(x) A(x,D_x)  v(x,\tau )\,dx \,d\tau\\
&  & 
  -\int_\varepsilon ^t \int_{{\mathbb R}^n} \tau ^{1-p} a(x)|u(x,\tau )|^p  v(x,\tau )\,dx \,d\tau \,.
\end{eqnarray*}
Further,
\begin{eqnarray*}
&    &
\int_\varepsilon ^t \int_{{\mathbb R}^n}  a(x)  u_{tt} (x,\tau )  v(x,\tau )\,dx \,d\tau  \\
& = &
  \int_{{\mathbb R}^n}   a(x) u_{t} (x,\tau )  v(x,\tau )\,dx \Big|_{\varepsilon }^t - \int_{{\mathbb R}^n}   a(x) u  (x,\tau )  v_t(x,\tau )\,dx \Big|_{\varepsilon }^t\\
  &  &
+ \int_\varepsilon ^t \int_{{\mathbb R}^n}   u  (x,\tau )  \tau ^{ - {4}/{3}} a(x)A(x,D_x)  v (x,\tau )\,dx \,d\tau \,.
\end{eqnarray*}
Hence,
\begin{eqnarray*}
&  &
 \int_{{\mathbb R}^n}  a(x)  u_{t} (x,\tau )  v(x,\tau )\,dx \Big|_{\varepsilon }^t - \int_{{\mathbb R}^n}   a(x) u (x,\tau )  v_t(x,\tau )\,dx \Big|_{\varepsilon }^t\\
&  =  &
 \int_\varepsilon ^t \int_{{\mathbb R}^n} \tau ^{1-p} a(x)|u(x,\tau )|^p  v(x,\tau )\,dx \,d\tau .
\end{eqnarray*}
The last equation implies
\begin{eqnarray*}
&  &
\left( \frac{d}{d\tau } \int_{{\mathbb R}^n}   a(x) u  (x,\tau )   v(x,\tau )\,dx   - 2\int_{{\mathbb R}^n}  a(x)  u  (x,\tau )  v_t(x,\tau )\,dx
\right)\Big|_{\varepsilon }^t \\
& = &
\int_\varepsilon ^t \int_{{\mathbb R}^n} \tau ^{1-p} a(x)|u(x,\tau )|^p  v(x,\tau )\,dx \,d\tau \,.
\end{eqnarray*}
It follows
\begin{eqnarray*}
&  &
 \frac{d}{d t} F_1(t)   - 2\frac{\lambda _t(t)}{\lambda  (t)} \int_{{\mathbb R}^n}   a(x) u  (x,t)  \lambda  (t) \varphi (x)\,dx  \\
& = &
 \frac{d}{d t} F_1(t)\Big|_{\varepsilon } - 2\frac{\lambda _t(\varepsilon )}{\lambda  (\varepsilon )}\int_{{\mathbb R}^n}  a(x)  u  (x,\varepsilon )  \lambda  (\varepsilon )\varphi (x)\,dx  \\
&  &
+ \int_\varepsilon ^t \int_{{\mathbb R}^n} \tau ^{1-p} a(x)|u(x,\tau )|^p  v(x,\tau )\,dx \,d\tau \,.
\end{eqnarray*}
Consequently,
\begin{eqnarray*}
 \frac{d}{d t} F_1(t)   - 2\frac{\lambda _t(t)}{\lambda  (t)} F_1(t)  
&  =  &  
 \frac{d}{d t} F_1(t)\Big|_{\varepsilon } - 2\frac{\lambda _t(\varepsilon )}{\lambda  (\varepsilon )}F_1(\varepsilon ) \\
&  &
+ \int_\varepsilon ^t \int_{{\mathbb R}^n} \tau ^{1-p} a(x)|u(x,\tau )|^p  v(x,\tau )\,dx \,d\tau \,.
\end{eqnarray*}
It follows
\begin{eqnarray*}
&  &
 \frac{d}{d t} \left(  F_1(t) \exp \left( -\int_\varepsilon ^t 2\frac{\lambda _t(\tau )}{\lambda  (\tau )} \,d \tau  \right)  \right)  \\
& = &
 \exp \left( -\int_\varepsilon ^t 2\frac{\lambda _t(\tau )}{\lambda  (\tau )} \,d \tau  \right)\\
&  &
\times  \left\{ \frac{d}{d t} F_1(t)\Big|_{\varepsilon } - 2\frac{\lambda _t(\varepsilon )}{\lambda  (\varepsilon )}F_1(\varepsilon ) + \int_\varepsilon ^t \int_{{\mathbb R}^n} \tau ^{1-p} a(x)|u(x,\tau )|^p  v(x,\tau )\,dx \,d\tau
 \right\}\,,
\end{eqnarray*}
that is
\begin{eqnarray*}
&  &
 \frac{d}{d t} \left(  F_1(t)    \left(\frac{\lambda (t )}{\lambda  (\varepsilon  )} \right) ^{-2}     \right)  \\
&  =  &
 \left(\frac{\lambda (t )}{\lambda  (\varepsilon  )} \right) ^{-2}     \left\{ \frac{d}{d t} F_1(t)\Big|_{\varepsilon } - 2\frac{\lambda _t(\varepsilon )}{\lambda  (\varepsilon )}F_1(\varepsilon ) + \int_\varepsilon ^t \int_{{\mathbb R}^n} \tau ^{1-p} a(x)|u(x,\tau )|^p  v(x,\tau )\,dx \,d\tau
 \right\}\,.
\end{eqnarray*}
We integrate it and obtain
\begin{eqnarray}
\label{6.10}
    F_1(t)
& = &
 \left(\frac{\lambda (t )}{\lambda  (\varepsilon  )} \right) ^{2} \Bigg[   F_1(\varepsilon )   +
\int_\varepsilon ^t  \left(\frac{\lambda (s )}{\lambda  (\varepsilon  )} \right) ^{-2}   \\
&  &
\times    \left\{ \frac{d}{d t} F_1(t)\Big|_{\varepsilon } - 2\frac{\lambda _t(\varepsilon )}{\lambda  (\varepsilon )}F_1(\varepsilon ) + \int_\varepsilon ^s \int_{{\mathbb R}^n} \tau ^{1-p} a(x)|u(x,\tau )|^p  v(x,\tau )\,dx \,d\tau
 \right\}\,ds \Bigg]\,. \nonumber
\end{eqnarray}
On the other hand, according to (iii) of Lemma~\ref{L6.2} we have
$
\displaystyle \frac{\lambda _t(t)}{\lambda (t)}    =  -\frac{3}{\sqrt[3]{t }(3\sqrt[3]{t }+1)}
$.
Consider the term
\begin{eqnarray*}
 \frac{d}{d t} F_1(t)\Big|_{\varepsilon } - 2\frac{\lambda _t(\varepsilon )}{\lambda  (\varepsilon )}F_1(\varepsilon ) 
 & = &
\int_{{\mathbb R}^n}  a(x)u_t(x,\varepsilon )\lambda (\varepsilon )\varphi (x)\,dx + \int_{{\mathbb R}^n}  a(x)u(x,\varepsilon )\lambda_t (\varepsilon )\varphi (x)\,dx\\
&  &
 +\frac{6}{\sqrt[3]{\varepsilon  }(3\sqrt[3]{\varepsilon  }+1)}  \int_{{\mathbb R}^n} a(x) u (x,\varepsilon )\lambda (\varepsilon )\varphi (x)\,dx.
 \end{eqnarray*}
We can rewrite it as follows
\begin{eqnarray*}
 &  &
 \frac{d}{d t} F_1(t)\Big|_{\varepsilon }
+\frac{6}{\sqrt[3]{\varepsilon  }(3\sqrt[3]{\varepsilon  }+1)}  F_1(\varepsilon ) \\
 & = &
 \int_{{\mathbb R}^n}  a(x) \left\{u_t(x,\varepsilon ) +3 \varepsilon^{-1/3} A(x,D_x)  \varphi_0 (x)\right\}v(x,\varepsilon )\,dx\\
 &  &
- \int_{{\mathbb R}^n}   3  a(x)\varepsilon^{-1/3} A(x,D_x)  \varphi_0 (x) v(x,\varepsilon )\,dx  \\
&  &
- \int_{{\mathbb R}^n} \frac{3}{\sqrt[3]{\varepsilon  }(3\sqrt[3]{\varepsilon  }+1)}  a(x)u(x,\varepsilon )v(x,\varepsilon )\,dx
 +\frac{6}{\sqrt[3]{\varepsilon  }(3\sqrt[3]{\varepsilon  }+1)}  \int_{{\mathbb R}^n} a(x) u (x,\varepsilon )v(x,\varepsilon )\,dx\\
& = &
 \int_{{\mathbb R}^n}   a(x)\{u_t(x,\varepsilon ) +3 \varepsilon^{-1/3} A(x,D_x)  \varphi_0 (x)\}v(x,\varepsilon )\,dx \\
 &  &
 + \int_{{\mathbb R}^n}  3 \varepsilon^{-1/3} a(x)\Big\{-   \varphi_0 (x)
 + \frac{1}{(3\sqrt[3]{\varepsilon  }+1)} u(x,\varepsilon )  \Big\} v(x,\varepsilon )\,dx\,.
 \end{eqnarray*}
Hence, taking into account the initial conditions for \, $u$, \, we derive
\begin{eqnarray*}
\lim_{\varepsilon \to 0^+} \left( \frac{d}{d t} F_1(t)\Big|_{\varepsilon }
- 2\frac{\lambda _t(\varepsilon )}{\lambda  (\varepsilon )}F_1(\varepsilon ) \right)
  =  
 \int_{{\mathbb R}^n}   a(x) \varphi_1 (x)\varphi (x)\,dx
- 9  \int_{{\mathbb R}^n}  a(x)\varphi_0 (x)
   \varphi (x)\,dx \,.
\end{eqnarray*}
Now
\begin{eqnarray*}
&  &
\lim_{\varepsilon \to 0^+}
 \left(\frac{\lambda (t )}{\lambda  (\varepsilon  )} \right) ^{2} \Bigg[   F_1(\varepsilon )
  +
\int_\varepsilon ^t  \left(\frac{\lambda (s )}{\lambda  (\varepsilon  )} \right) ^{-2}     \left\{ \frac{d}{d t} F_1(t)\Big|_{\varepsilon } - 2\frac{\lambda _t(\varepsilon )}{\lambda  (\varepsilon )}F_1(\varepsilon ) \right\} ds \Bigg] \\
& = &
 \lambda (t )^{2}   F_1(0)
  +
\lim_{\varepsilon \to 0^+} \left(\frac{\lambda (t )}{\lambda  (\varepsilon  )} \right) ^{2}  \int_\varepsilon ^t  \left(\frac{\lambda (s )}{\lambda  (\varepsilon  )}  \right) ^{-2}     \left\{ \frac{d}{d t} F_1(t)\Big|_{\varepsilon } - 2\frac{\lambda _t(\varepsilon )}{\lambda  (\varepsilon )}F_1(\varepsilon ) \right\}  ds \\
& = &
 \lambda (t )^{2}   F_1(0)
  +
 \left\{ \int_{{\mathbb R}^n}  a(x)  \varphi_1 (x)\varphi (x)\,dx
- 9  \int_{{\mathbb R}^n} a(x) \varphi_0 (x)
   \varphi (x)\,dx\right\} \lambda^{2} (t )   \int_0 ^t    \lambda^{-2} (s )           ds \\
& = &
(3\sqrt[3]{t}+1)^{2}\exp \left( -6\sqrt[3]{t}  \right)    \int_{{\mathbb R}^n} \varphi_0 (x)
   \varphi (x)\,dx \\
&  &
\hspace{1.5cm}   +  \left\{ \int_{{\mathbb R}^n}  a(x)  \varphi_1 (x)\varphi (x)\,dx
- 9  \int_{{\mathbb R}^n}  a(x)\varphi_0 (x)
   \varphi (x)\,dx\right\} \\
&  &
\hspace{3cm} \times (3\sqrt[3]{t}+1)^{2}\exp \left( -6\sqrt[3]{t}  \right)   \int_0 ^t   (3\sqrt[3]{s}+1)^{-2}\exp \left( 6\sqrt[3]{s}  \right)    ds \,.
\end{eqnarray*}
On the other hand
\begin{eqnarray*} 
\int_0 ^t   (3\sqrt[3]{s}+1)^{-2}\exp \left( 6\sqrt[3]{s}  \right)    ds =
\frac{1}{18} \left(\exp \left( 6\sqrt[3]{t}  \right)\frac{   3 \sqrt[3]{t}-1 }{3 \sqrt[3]{t}+1}+1\right)
\end{eqnarray*}
implies
\begin{eqnarray*}
&  &
\lim_{\varepsilon \to 0^+}
 \left(\frac{\lambda (t )}{\lambda  (\varepsilon  )} \right) ^{2} \Bigg[   F_1(\varepsilon )
  +
\int_\varepsilon ^t  \left(\frac{\lambda (s )}{\lambda  (\varepsilon  )} \right) ^{-2}     \left\{ \frac{d}{d t} F_1(t)\Big|_{\varepsilon } - 2\frac{\lambda _t(\varepsilon )}{\lambda  (\varepsilon )}F_1(\varepsilon ) \right\}\Bigg] ds \\
& = &
(3\sqrt[3]{t}+1)^{2}\exp \left( -6\sqrt[3]{t}  \right)    \int_{{\mathbb R}^n}  a(x)\varphi_0 (x)
   \varphi (x)\,dx \\ 
&  & 
\hspace{1.5cm}  +
 \left\{ \int_{{\mathbb R}^n}   a(x) \varphi_1 (x)\varphi (x)\,dx
- 9  \int_{{\mathbb R}^n} \varphi_0 (x)
   \varphi (x)\,dx\right\} \\
&  &
\hspace{3.5cm} \times (3\sqrt[3]{t}+1)^{2}  \frac{1}{18} \left(\frac{   3 \sqrt[3]{t}-1 }{3 \sqrt[3]{t}+1}+\exp \left( -6\sqrt[3]{t}  \right) \right)\,.
\end{eqnarray*}
Due to the conditions of the lemma,
\begin{eqnarray*}
    \int_{{\mathbb R}^n}    a(x)\varphi_1 (x) \varphi (x)\,dx
- 9  \int_{{\mathbb R}^n}  a(x)\varphi_0 (x)
 \varphi (x)\,dx
\geq   \frac{1}{2}\int_{{\mathbb R}^n} a(x) \varphi_1 (x)
 \varphi (x)\,dx  >0\,.
\end{eqnarray*}
Then, from (\ref{6.10}), by letting $\varepsilon \to 0 $, we derive
\begin{eqnarray*}
    F_1(t)
& \geq  &
 \left(\frac{\lambda (t )}{\lambda  (\varepsilon  )} \right) ^{2} \Bigg[   F_1(\varepsilon )
  +
\int_\varepsilon ^t  \left(\frac{\lambda (s )}{\lambda  (\varepsilon  )} \right) ^{-2}     \left\{ \frac{d}{d t} F_1(t)\Big|_{\varepsilon } - 2\frac{\lambda _t(\varepsilon )}{\lambda  (\varepsilon )}F_1(\varepsilon ) +
 \right\}\,ds \Bigg]\\
& \geq  &
(3\sqrt[3]{t}+1)^{2}\exp \left( -6\sqrt[3]{t}  \right)    \int_{{\mathbb R}^n}  a(x)\varphi_0 (x)
   \varphi (x)\,dx \\
&  &
  +   (3\sqrt[3]{t}+1)^{2}  \frac{1}{18} \left(\frac{   3 \sqrt[3]{t}-1 }{3 \sqrt[3]{t}+1}+\exp \left( -6\sqrt[3]{t}  \right) \right) \frac{1}{2} \int_{{\mathbb R}^n}   a(x) \varphi_1 (x)\varphi (x)\,dx  \\
& \geq  &
(3\sqrt[3]{t}+1)^{2}\exp \left( -6\sqrt[3]{t}  \right)    \int_{{\mathbb R}^n}  a(x)\varphi_0 (x)
   \varphi (x)\,dx \\
   &  &
  +   (9\sqrt[3]{t^2}-1)   \frac{1}{36}    \int_{{\mathbb R}^n}  a(x)  \varphi_1 (x)\varphi (x)\,dx \,.
\end{eqnarray*}
Lemma is proved. \hfill $\square$
\medskip

Furthermore,  the inequality (\ref{3.12})  implies
\begin{eqnarray*}
F''(t)
& \geq  &
\lambda ^{-p} (t)\left( \int_{|x| \le R+  \phi (t)} |\varphi (x )|^{p/(p-1)} \,dx \right)^{1- p}
t^{1-p}\left| F_1(t) \right|^{p}\quad \mbox{\rm for all}\quad t \geq R_A+1\,.
\end{eqnarray*}
According to the last lemma 
\begin{eqnarray*}
&  &
F''(t) \\
& \geq  & 
c_R \lambda ^{-p} (t)   (R+   \phi (t))^{ -\frac{n-1}{2} (p-2)  }e^{ - \phi (t) p}
t^{1-p}\left| F_1(t) \right|^{p}\\
& \geq  &
c_R    (R+   \phi (t))^{-p -\frac{n-1}{2} (p-2)  }
t^{1-p} \left|(9t^{\frac{2}{3}}-1)      \int_{{\mathbb R}^n}   a(x) \varphi_1 (x)\varphi (x)\,dx \right|^{p} \quad \mbox{\rm for all}\quad t \geq R_A+1\,.
\end{eqnarray*}
Finally 
\begin{equation}
\label{6.13}
F''(t)   \geq  
C_R    (R+  \phi (t))^{-p -\frac{n-1}{2} (p-2)  }
t^{1-p+\frac{2}{3}p}    \left|
\int_{{\mathbb R}^n}  a(x)\varphi_1 (x)
 \varphi (x)\,dx    \right|^{p}\,\, \mbox{\rm for all} \, t \geq R_A+1\,.
\end{equation}
For $t > 1$ and arbitrary $\varepsilon \in (0,1) $,    it follows
 \begin{eqnarray*}
F(t)
& = &
F(\varepsilon )+\int_\varepsilon ^1 \left\{ F'(\varepsilon ) + \int_\varepsilon ^{t_1} F'{}'(t_2) \, dt_2 \right\} \, dt_1
+\int_1^t \left\{ F'(\varepsilon ) + \int_\varepsilon ^{t_1} F'{}'(t_2) \, dt_2 \right\} \, dt_1\\
& \geq  &
F(\varepsilon )+\int_\varepsilon ^1   F'(\varepsilon )  \, dt_1
+\int_1^t   F'(\varepsilon )   \, dt_1
+\int_1^t \left\{\int_1^{t_1} F'{}'(t_2) \, dt_2 \right\} \, dt_1\\
& \geq  &
F(\varepsilon )
+  (t-\varepsilon ) F'(\varepsilon )
+\int_1^t \left\{\int_1^{t_1} F'{}'(t_2) \, dt_2 \right\} \, dt_1\,.
 \end{eqnarray*}
 By letting $\varepsilon \to 0 $ and using (\ref{6.13}) we derive
\begin{eqnarray*} 
F(t)
& \geq  &
tF'(0)+ F(0)\\
&  &
+\,
c_R  \left|
\int_{{\mathbb R}^n}  a(x)\varphi_1 (x)
 \varphi (x)\,dx   \right|^{p} \int_1^t \int_1^{t_2} (R+   \phi (t_1))^{-p -\frac{n-1}{2} (p-2)  }
t_1^{1-\frac{1}{3}p}
 dt_1 \, dt_2  \,.
\end{eqnarray*}
Set (see (\ref{fdtdt}))
\[ 
r   =  
\frac{1}{6} \left[2n+16-(n+3)p   \right] \,, \qquad 
q  =  
 \frac{(n+3)(p-1)}{3}\,.
\]
We need $r\geq 1$ that is, $p\leq  (2n+10)/(n+3)$. 
The Kato's lemma  (see, e.g., \cite[Lemma~2.1]{Yag_2005}), concerning differential inequalities 
\begin{eqnarray*}
F(t)
& \geq  &
c_0   (1+t)^r \quad \mbox{for large }\quad t,\\
F '' (t)
& \geq &
   (1+t)^{ -q}|F(t)|^p  \quad \mbox{for large }\quad t\,,
\end{eqnarray*} 
conditions are $r\geq 1 $, $p>1$ and
\begin{eqnarray*}
&  &
(p-1)r>q-2\Longleftrightarrow 
   (n+3)p^2   -(n+13)p   -2 <0  \,.
\end{eqnarray*}
Due to the definition of   $p_{cr}(n) $  we obtain $p< p_{cr}(n) $. 
   The theorem is proved. \hfill $\square$

\begin{corollary}
For the covariant semilinear wave equation with $n=3$ and $F(\psi )=|\psi |^p$ assume that $  1< p<3$.
Then for every arbitrary small number $\varepsilon >0 $ and an arbitrary number  $s$ there exist functions  $\varphi_0 , \varphi_1  \in C_0^\infty({\mathbb  R}^3) $, supp\,$\varphi_0 , \varphi_1 \subseteq \{x \in {\mathbb  R}^3\,|\, |x| \leq R \} $
with norms satisfying  inequality
\[
\|\varphi_0\|_{H_{(s)}({\mathbb  R}^3)}+ \|\varphi_1\|_{H_{(s)}({\mathbb  R}^3)} <\varepsilon
\]
such that the solution of the problem
(\ref{WEEDS_A}) with support in $\{(x,t) \,|\, t>0, \,\,x \in B_{R+\phi (t)s_A }(0) \}$  blows up in finite time.
\end{corollary}

Now  we analyze  the conditions of the theorem. 
From the graph it follows that for $n\leq 4$ there is a small data  blowing up solution if $1<p<1+\frac{6}{n}$. For  the dimensions $n \geq 5$ such solution exists if $1<p <  \frac{n+13+\sqrt{n^2+34 n+193}}{2 (n+3)}  $.    
\begin{center}
\begin{figure}[ht] 
\includegraphics[width=0.29\textwidth]{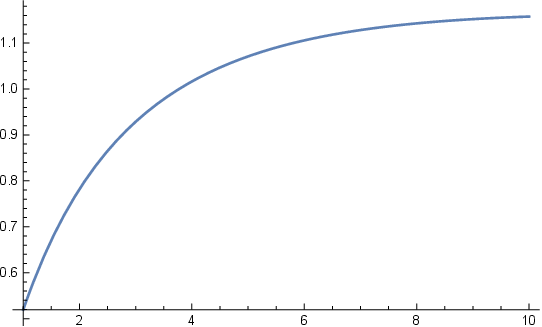} \hspace{0.4cm} 
\includegraphics[width=0.29\textwidth]{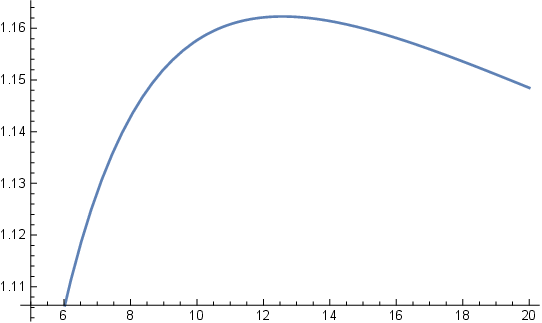} \hspace{0.4cm} 
\includegraphics[width=0.29\textwidth]{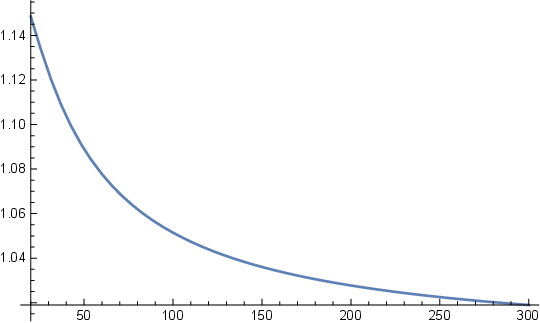} 
\caption{ \small 
$ { \frac{n+13+\sqrt{n^2+34 n+193}}{2 (n+3)} }/({1+\frac{6}{n}}) $,   
\hspace{0.4cm}  $n \in [1,10]$ \hspace{0.4cm}  $n \in [4,20]$ \,\hspace{0.4cm} \small  $n \in [20,300]$  }
\end{figure}
\end{center}

\section{Equation without singularity. Proof of Theorem~\ref{T6.9}}
\setcounter{equation}{0}
\renewcommand{\theequation}{\thesection.\arabic{equation}}
\label{S3}

Theorem~\ref{T6.9} shows that  the blow-up, which is stated in Theorem~\ref{T3.2},  is caused by the semilinear term.  Consider the   Cauchy problem (\ref{6.16_intr}). 
Let $p_{0}(n,k) $ be a positive root of the equation (\ref{pcrII}). 
\begin{figure}[ht]
\begin{center}
\includegraphics[width=0.4\textwidth]{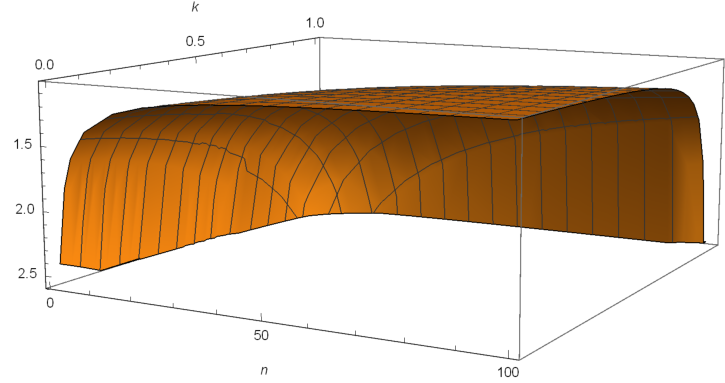}  
\caption{Function $p_0(n,k)$ for $0\leq k\leq 1 $ and $1\leq n\leq 100 $}
\end{center}
\end{figure}
(For the graph $p=p_{0}(n,k) $ see Figure~2.) The equation  of (\ref{6.16_intr}) is strictly hyperbolic for every bounded  interval of time and it has smooth coefficients. Consequently, for every smooth initial functions $\varphi_0  $ and $ \varphi_1 $ the problem (\ref{6.16_intr}) has a local solution.  According to Theorem~\ref{T6.9}  for $n=3$ and $p <  3$  a local in time solution, in general, cannot be prolonged to the global solution.    
\medskip

\noindent
{\bf Proof of Theorem~\ref{T6.9}.} We use operators ${\mathcal L}  $ and ${\mathcal S} $ which are  introduced above in (\ref{OPLS}):
$
{\mathcal L} :=\partial_{t}^2
-  t^{-2k} A(x,D_x)      + 2 t^{-1 }        \partial_t  $, $
{\mathcal S}:= \partial_{t}^2
-  t^{-2k} A(x,D_x)    \,,
$
and for $t \not= 0$ the operator identity (\ref{OPI}). 
The last identity suggests  a change of unknown function $\psi $ with $u$ such that
$
\psi =t^{-1}u $.
 The problem for $u$ is as follows:
\begin{eqnarray}
\label{2.2ab}
\begin{cases}
 u_{tt} - t^{-2k} A(x,D_x)     u  = t^{1-p} |u|^p,  \qquad t>1 ,\,\, x \in {\mathbb R}^n,\cr
 u (x,1) = u_0 (x)\,,\quad  u_0 (x):=\varphi_0 (x) , \qquad  \,\, x \in {\mathbb R}^n, \cr
u _t (x,1)=  u _1 (x)\,,\quad  u_1 (x):=\varphi_0 (x) + \varphi_1 (x),   \quad x \in {\mathbb  R}^n .
\end{cases}
\end{eqnarray}
Denote
\[
F (t)= \int_{{\mathbb R}^{n}}a(x) u(x,t) \,dx \, .
\]
Then $F\in C^2[1,T] $, provided that the function $u$ is defined for all $(x,t) \in {\mathbb R}^n\times [1,T] $, and
\begin{eqnarray*} 
 \displaystyle     F(1)=  \int_{{\mathbb R}^{n}}a(x)  u_0 (x)  \,dx =C_0 >0,\quad F'(1)
  =    
 \int_{{\mathbb R}^{n}}  \,a(x)  u_1 (x)\,dx =C_1\,.  
\end{eqnarray*}
From the  equation of (\ref{2.2ab}) we have
\begin{eqnarray}
\label{3.22new}
F ''(t) 
& = &
t^{1-p} \int_{{\mathbb R}^{n}} a(x) |u(x,t)|^p \,dx \geq 0 \quad \mbox{for all}\quad t>1.
\end{eqnarray}
Furthermore,
\begin{eqnarray*}
F (t)
&= &
F (1 )
+ (t-1 ) F' (1 )  +\int_1^t   \int_1 ^{t_1} F'{}'(t_2 )d t_2  \, d t_1  \\
&\geq  &
F (1 )
+ (t-1 ) F' (1 )    \geq 0 \quad \mbox{for all}\quad t\geq 1\,,
\end{eqnarray*}
provided that $C_0\geq 0$ and $C_1\geq 0$. Hence
\begin{eqnarray*}
F (t)
& \geq  &
(t-1 ) \int_{{\mathbb R}^{n}} u_1 (x) \,dx+\int_{{\mathbb R}^{n}} u _0 (x)  \,dx \geq 0 \quad \mbox{for all}\quad t\geq 1\,.
\end{eqnarray*}
Assume that supp$\,u_i \subseteq B_R(0) $, $i=0,1$. On the other hand,
using the compact support of $u(\cdot,t)$ and H\"older's inequality  
with \,   $\phi (t):=\frac{1}{1-k}t^{1-k} $ we obtain
\begin{eqnarray*}
\left| \int_{{\mathbb R}^n} a(x) u (x,t) \,dx \right|^{p}
& \le &
\left( \int_{|x| \le R+ \phi (t) -\phi (1)} 1 \,dx \right)^{ p-1 }
\left( \int_{|x| \le R+ \phi (t)-\phi (1)} a(x) ^p |u (x,t)|^{p} \,dx \right)\\
& \lesssim  &
(R+\phi (t))^{ n(p-1)}
\left( \int_{|x| \le R+ \phi (t)-\phi (1)}a(x)  |u (x,t)|^{p} \,dx \right) \,.
\end{eqnarray*}
Hence from (\ref{3.22new}) we derive
\begin{eqnarray*} 
F '' (t)
& \geq &
    (1+t)^{1-p-n(p-1)(1-k) } |F(t)|^p\quad \mbox{for all}\quad t\geq 1 \,.
\end{eqnarray*}
We  set
\begin{eqnarray*}
r=1 \,,\qquad q:= (p-1)+n(p-1)(1-k)=   (p-1)(1+n (1-k))\,.
\end{eqnarray*}
 Consider the first case of $1<p<1+\frac{2}{n (1-k)}$.  If $1< p<1 +\frac{2}{n (1-k)}$  and $C_1>0$,  then we can apply Kato's lemma (see, e.g., \cite[Lemma~2.1]{Yag_2005}) since
\[
p-1>(p-1)(1+n (1-k))-2\Longleftrightarrow p<1+ \frac{2}{n (1-k)} \,.
\]
Thus, the solution blows up. \\
\medskip

 Consider the second case.  For this case we  
choose
\begin{eqnarray*}
v(x,t)
& := &
\widetilde{\lambda }(t)\varphi (x)\,,\qquad 
\widetilde{\lambda }(t)
  :=   
\frac{1}{K_{\frac{1}{2 -2k}}\left(\phi (1) \right)} \sqrt{t} K_{\frac{1}{2 -2k}}\left(\phi (t) \right)\,,
\end{eqnarray*}
where $K_a(z) $ is the modified Bessel function of the second kind. The function $\widetilde{\lambda }=\widetilde{\lambda }(t)$ solves the  equation
\[
\widetilde{\lambda }_{tt}- t^{-2k}\widetilde{\lambda } =0\,.
\]
It is easy to verify the following limit
\[
\lim_{t \to \infty}\sqrt{t} K_{\frac{1}{2 -2k}}\left(\phi (t) \right)=0\,.
\]
Hence
\begin{eqnarray*}
v(x,1)
  = \varphi (x)\,,\qquad
\lim_{t \to \infty} v(x,t) =0  \,.
\end{eqnarray*}
We skip the  proof of  the next lemma.
\begin{lemma} 
\label{L6.13}
There is a number $\Lambda _0 > 0$ such that
\begin{eqnarray*}
\Lambda _1(k) := -  \widetilde{\lambda} _t(1)  
 = \frac{K_{\frac{1-2 k}{2-2 k}}\left(\frac{1}{1-k}\right)}{K_{\frac{1}{2-2 k}}\left(\frac{1}{1-k}\right)} > \Lambda_0  \quad \mbox{for all} \quad k \in[0,1)  \,.
\end{eqnarray*}
\end{lemma}
Assume that  $ u_0, u_1 \in C_0^\infty$, $\mbox{\rm supp\,}u_0, \mbox{\rm supp\,}u_1\subseteq \{ x \in {\mathbb R}^n\, |\, |x| \le R\}$. Now we turn to the function 
\[
F_1(t):=  \int_{{\mathbb R}^n}a(x)  u (x,t) v(x,t) \,dx 
\]
and obtain
\begin{eqnarray*}
\left| F_1(t) \right|^{p}
& \lesssim  &
\left( \int_{|x| \le R+ \phi (t)-\phi (1)} |v(x,t)|^{p/(p-1)} \,dx \right)^{ p-1 }
t^{p-1}F''(t)\quad \mbox{for all} \quad t>1\,.
\end{eqnarray*}
The  last estimate implies
\begin{eqnarray}
\label{3.12a}
F''(t)
& \geq  &
\left( \int_{|x| \le R+ \phi (t)-\phi (1)} |v(x,t)|^{p/(p-1)} \,dx \right)^{1- p }
t^{1-p}\left| F_1(t) \right|^{p}\quad \mbox{for all} \quad t>1\,.
\end{eqnarray}
\begin{lemma}
\label{L3.2}
Assume that  $ u_0, u_1 \in C_0^\infty$, $\mbox{\rm supp\,}u_0, \mbox{\rm supp\,}u_1\subseteq B_R(0) \subseteq   {\mathbb R}^n $, 
and
\begin{eqnarray*}
  \Lambda _1(k)\int_{{\mathbb R}^n} a (x) u_0 (x)\varphi (x)dx+ \int_{{\mathbb R}^n} a (x) u_1 (x)\varphi (x)dx \geq c_0 \int_{{\mathbb R}^n} a (x) u_0 (x)\varphi (x)dx >  0\,.
\end{eqnarray*}

Then there exists a sufficiently large $T>1$ such that for the solution $u=u(x,t) $ of the problem (\ref{2.2ab}) with the support in $\{ x \in {\mathbb R}^n\, |\, |x| \le R+ \phi (t)-\phi (1) \}  $  one has
\[
  F_1(t)
\geq
\frac{1}{16}  t^k   \left\{ \Lambda _1(k)\int_{{\mathbb R}^n} a (x) u_0 (x)\varphi (x)dx+ \int_{{\mathbb R}^n} a (x) u_1 (x)\varphi (x)dx\right\}
 \quad \mbox{  for all }\quad t >T\,.
\]
\end{lemma}
\medskip

\noindent
{\bf Proof.} We have
\[
F_1(1) = \int_{{\mathbb R}^n} a (x) u (x,1) v(x,1)\,dx =   \int_{{\mathbb R}^n} a (x) u _0 (x) \varphi (x) \,dx \geq c_0 >0
\]
and
\begin{eqnarray*}
0
& = &
\int_1 ^t \int_{{\mathbb R}^n} a (x) ( u_{tt} (x,\tau ) -\tau ^{ - 2k}A(x,D_x) u -\tau ^{1-p}|u|^p) v(x,\tau )\,dx \,d\tau  \\
& =  &
\int_1 ^t \int_{{\mathbb R}^n} a (x)   u_{tt} (x,\tau )   v(x,\tau )\,dx \,d\tau  
   -\int_1 ^t \int_{{\mathbb R}^n}   \tau ^{ - 2k}a (x) u  A(x,D_x) v(x,\tau )\,dx \,d\tau\\
&  &
  -\int_1 ^t \int_{{\mathbb R}^n} \tau ^{1-p}a (x) |u|^p  v(x,\tau )\,dx \,d\tau \,.
\end{eqnarray*}
Further,
\begin{eqnarray*}
&    &
\int_1^t \int_{{\mathbb R}^n}  a (x)  u_{tt} (x,\tau )  v(x,\tau )\,dx \,d\tau  \\
& = &
  \int_{{\mathbb R}^n} a (x)   u_{t} (x,\tau )  v(x,\tau )\,dx \Big|_{\tau =1 }^{\tau =t} - \int_{{\mathbb R}^n}  a (x)  u  (x,\tau )  v_t(x,\tau )\,dx \Big|_{\tau =1 }^{\tau =t}\\
  &  &
+ \int_1 ^t \int_{{\mathbb R}^n}  u  (x,\tau )  t^{ - 2k} a (x) A(x,D_x) v (x,\tau )\,dx \,d\tau \,.
\end{eqnarray*}
Hence,
\begin{eqnarray*}
&  &
 \int_{{\mathbb R}^n} a(x)  u_{t} (x,\tau )  v(x,\tau )\,dx \Big|_{\tau =1 }^{\tau =t}
- \int_{{\mathbb R}^n}   a(x)  u  (x,\tau )   v_t(x,\tau )\,dx \Big|_{\tau =1 }^{\tau =t}\\
& = &
 \int_1 ^t \int_{{\mathbb R}^n} \tau ^{1-p} a(x) |u(x,\tau )|^p  v(x,\tau )\,dx \,d\tau
\end{eqnarray*}
implies
\begin{eqnarray*}
&  &
\left( \frac{d}{d \tau } \int_{{\mathbb R}^n}   a(x)  u  (x,\tau )   v(x,\tau )\,dx   - 2\int_{{\mathbb R}^n}    a(x) u  (x,\tau )  v_t(x,\tau )\,dx
\right)\Big|_{\tau =1 }^{ \tau =t} \\
& = &
\int_1 ^t \int_{{\mathbb R}^n} \tau ^{1-p} a(x) |u(x,\tau )|^p  v(x,\tau )\,dx \,d\tau
\end{eqnarray*}
and
\begin{eqnarray*}
&  &
 \frac{d}{d t} F_1(t)   - 2\frac{\widetilde{\lambda} _t(t)}{\widetilde{\lambda} (t)} \int_{{\mathbb R}^n}   a(x)  u  (x,t) \widetilde{\lambda} (t) \varphi (x)\,dx  \\
& = &
 \frac{d}{d t} F_1(t)\Big|_{1 } - 2\frac{\widetilde{\lambda} _t(1 )}{\widetilde{\lambda} (1 )}\int_{{\mathbb R}^n}  a(x)   u  (x,1 )  \widetilde{\lambda}  (1 )\varphi (x)\,dx  \\
&  &
+ \int_1^t \int_{{\mathbb R}^n} \tau ^{1-p} a(x) |u(x,\tau )|^p  v(x,\tau )\,dx \,d\tau \,.
\end{eqnarray*}
On the other hand,
\begin{eqnarray*}
 \frac{\widetilde{\lambda} _t(t)}{\widetilde{\lambda} (t)}
& = &
-\frac{t^{-k} K_{\frac{1-2 k}{2-2 k}}\left(\phi (t)\right)}{K_{\frac{1}{2-2 k}}\left(\phi (t)\right)}<0
\quad \mbox{for all} \quad t>0\,,\\
\lim_{t \to \infty}
 \frac{\widetilde{\lambda} _t(t)}{\widetilde{\lambda} (t)}
& = &
0\,,\quad  \frac{\widetilde{\lambda} _t(1)}{\widetilde{\lambda} (1)}
 = \widetilde{\lambda} _t(1)
 = -\frac{K_{\frac{1-2 k}{2-2 k}}\left(\frac{1}{1-k}\right)}{K_{\frac{1}{2-2 k}}\left(\frac{1}{1-k}\right)}\,.
\end{eqnarray*}
Consequently,
\begin{eqnarray*}
&  &
 \frac{d}{d t} F_1(t)   - 2\frac{\widetilde{\lambda} _t(t)}{\widetilde{\lambda} (t)} F_1(t)  \\
& = &
 \frac{d}{d t} F_1(t)\Big|_{1 } - 2\frac{\widetilde{\lambda} _t(1 )}{\widetilde{\lambda} (1 )}F_1(1)  + \int_1^t \int_{{\mathbb R}^n} \tau ^{1-p} a(x) |u(x,\tau )|^p  v(x,\tau )\,dx \,d\tau \,,
\end{eqnarray*}
that is, 
\begin{eqnarray*}
&  &
 \frac{d}{d t} \left(  F_1(t)    \left(\widetilde{\lambda}(t ) \right) ^{-2}     \right) \\
& = &
 \left(\widetilde{\lambda}(t ) \right) ^{-2}     \left\{ \frac{d}{d t} F_1(t)\Big|_{1 } + 2\Lambda_1 (k) F_1(1 ) + \int_1 ^t \int_{{\mathbb R}^n} \tau ^{1-p} a(x) |u(x,\tau )|^p  v(x,\tau )\,dx \,d\tau  \right\}\,,
\end{eqnarray*}
where, due to Lemma~\ref{L6.13},   $ 
\Lambda_1 (k) = - \widetilde{\lambda} _t(1 ) =\frac{K_{\frac{1-2 k}{2-2 k}}\left(\frac{1}{1-k}\right)}{K_{\frac{1}{2-2 k}}\left(\frac{1}{1-k}\right)}>\Lambda _0>0$\,. 
We integrate the last relation  
\begin{eqnarray*} 
    F_1(t)    \left( \widetilde{\lambda}(t )  \right) ^{-2}    
& = &
   F_1(1 )    +
\int_1 ^t   \left( \widetilde{\lambda}(s )  \right) ^{-2}   \Bigg\{ \frac{d}{d t} F_1(t)\Big|_{t=1 }   \\
&  &
+2\Lambda_1 (k)F_1(1 ) + \int_1^s \int_{{\mathbb R}^n} \tau ^{1-p}  a(x) |u(x,\tau )|^p  v(x,\tau )\,dx \,d\tau
 \Bigg\}\,ds\,.
\end{eqnarray*}
Finally,
\begin{eqnarray*} 
    F_1(t)  
& = &
 \left( \widetilde{\lambda}(t )  \right)^{2} \Bigg[   F_1(1 )   +
\int_1 ^t   \left( \widetilde{\lambda}(s )  \right) ^{-2}     \\
&  &
\times  \left\{ \frac{d}{d t} F_1(t)\Big|_{t=1 } +2\Lambda_1 (k)F_1(1 ) + \int_1^s \int_{{\mathbb R}^n} \tau ^{1-p} a(x) |u(x,\tau )|^p  v(x,\tau )\,dx \,d\tau
 \right\}\,ds \Bigg]\,.
\end{eqnarray*}
Consider two first terms of the integrand
\begin{eqnarray*} 
&  &
\frac{d}{d t} F_1(t)\Big|_{t=1 } +2\Lambda_1F_1(1 ) \\
& = &
 \int_{{\mathbb R}^n}  a(x) u_0 (x)v_t(x,1)dx+ \int_{{\mathbb R}^n}  a(x) u_1 (x)v (x,1)dx+ 2\Lambda_1\int_{{\mathbb R}^n}  a(x) u_0 (x)v (x,1)dx \\
& = &
\Lambda _1 (k)\int_{{\mathbb R}^n}  a(x) u_0 (x)   \varphi  (x )\, dx+ \int_{{\mathbb R}^n}  a(x) u_1 (x)\varphi  (x )\, dx \,.
\end{eqnarray*}
Then
\begin{eqnarray}
\label{3.19b}
&  &
 \left( \widetilde{\lambda}(t )  \right)^{2} \Bigg[   F_1(1 )
  +
\int_1 ^t   \left( \widetilde{\lambda}(s )  \right) ^{-2}     \left\{ \frac{d}{d t} F_1(t)\Big|_{t=1 } +2 \Lambda _1 (k)F_1(1 ) \right\}\,ds \Bigg]\\
& = &
 \left( \widetilde{\lambda}(t )  \right)^{2}   F_1(1 ) \nonumber \\
&  &
  +
 \left[ \Lambda _1 (k)\int_{{\mathbb R}^n}  a(x) u_0 (x)\varphi (x)dx 
+ \int_{{\mathbb R}^n}  a(x) u_1 (x)\varphi (x)dx\right]  \left( \widetilde{\lambda}(t )  \right)^{2} \int_1 ^t   \left( \widetilde{\lambda}(s )  \right) ^{-2} \,ds   \nonumber \,.
\end{eqnarray} 
The following lemma completes the proof of Lemma~\ref{L3.2}. \hfill $\square$ 
\begin{lemma} 
\label{L6.14b}
There is a number $T_1>0$ such that
\begin{eqnarray*}  
 \widetilde{\lambda}  ^{2}(t ) \int_T ^t    \widetilde{\lambda}   ^{-2}(s ) \,ds   
& \geq &
 \frac{1}{32} t^{k }  
\quad \mbox{for all}\quad t\geq  T_1 \,.
\end{eqnarray*}
\end{lemma} 
\medskip

\noindent
{\bf Proof.} For all  $T>1$ we have
\begin{eqnarray}
\label{6.30} 
  \widetilde{\lambda}  ^{2}(t ) \int_1 ^t    \widetilde{\lambda}   ^{-2}(s ) \,ds    =   \widetilde{\lambda}  ^{2}(t ) \int_1 ^T    \widetilde{\lambda}   ^{-2}(s ) \,ds  
+  \widetilde{\lambda}  ^{2}(t ) \int_T ^t    \widetilde{\lambda}   ^{-2}(s ) \,ds  
\quad \mbox{for all}\quad t\geq T\,.
\end{eqnarray}
For  large $t$ there is the following asymptotic 
\begin{eqnarray*} 
\sqrt{t} K_{\frac{1}{2-2 k}}\left(\phi (t)\right)
= \sqrt{\frac{\pi }{2}} \sqrt{1- k} e^{-\phi (t)} t^{k/2}  \left( 1+o(1 )\right)\,.
\end{eqnarray*}
Consider the second integral; for the sufficiently large $T$ we have 
\begin{eqnarray*} 
&  &
 \widetilde{\lambda}  ^{2}(t ) \int_T ^t    \widetilde{\lambda}   ^{-2}(s ) \,ds  \\
& \geq &
\frac{1}{2}  e^{-2\frac{t^{1-k}}{1-k}} t^{2k } \int_T ^t    e^{2\frac{s^{1-k}}{1-k}}s^{-2k }  \,ds \\
& = &
 \frac{1}{2} e^{-2\frac{t^{1-k}}{1-k}} t^{2k }\frac{1}{4} 
\Bigg(2 e^{\frac{2 t^{1-k}}{1-k }} t^{-k}+ k e^{ \frac{2 t^{1-k}}{1-k }}  t^{-1} +2^{\frac{1}{1-k}} k \left(\frac{1}{k-1}\right)^{\frac{k-2}{k-1}} \Gamma \left(\frac{1}{k-1},\frac{2 t^{1-k}}{k-1}\right)\\
&  &
-2 e^{-\frac{2 T^{1-k}}{k-1}} T^{-k}-\frac{k e^{-\frac{2 T^{1-k}}{k-1}}}{T}-2^{\frac{1}{1-k}} k \left(\frac{1}{k-1}\right)^{\frac{k-2}{k-1}} \Gamma \left(\frac{1}{k-1},\frac{2 T^{1-k}}{k-1}\right)\Bigg)
\quad \mbox{for all}\quad t\geq T \,,
\end{eqnarray*}
where \, $\Gamma (a,z)= \int_z^\infty t^{a-1}e^{-t}\,dt $\, is the incomplete gamma function. (See, e.g., \cite[Sec.6.9.2]{B-E}.)
On the other hand, since $k=1-\varepsilon  $, $\varepsilon >0 $, we obtain for the incomplete gamma function the following 
asymptotic formula (see \cite[Sec.6.13.1]{B-E})
\begin{eqnarray*}
\left(\frac{1}{k-1}\right)^{\frac{k-2}{k-1}} \Gamma \left(\frac{1}{k-1},\frac{2 t^{1-k}}{k-1}\right)
& = &
2^{\frac{3-2 k}{k-1}}   e^{-\frac{2 t^{1-k}}{k-1}} t^{k-2} \left(2  +O(t^{k-1})\right)\\
& \leq  &
ce^{-\frac{2 t^{1-k}}{k-1}}t^{-1-\varepsilon }
\quad \mbox{for all}\quad t\geq T \,.
\end{eqnarray*}
Consequently, for the sufficiently large $T_1>T$ we obtain
\begin{eqnarray*}  
 \widetilde{\lambda}  ^{2}(t ) \int_T ^t    \widetilde{\lambda}   ^{-2}(s ) \,ds  
& \geq &
 \frac{1}{2} e^{-2\frac{t^{1-k}}{1-k}} t^{2k }\frac{1}{4} 
\Bigg(  e^{\frac{2 t^{1-k}}{1-k }} t^{-k}
-2 e^{\frac{2 T^{1-k}}{1-k}} T^{-k}\\
&  &
\hspace{2cm}  -\frac{k e^{-\frac{2 T^{1-k}}{k-1}}}{T}-2^{\frac{1}{1-k}} k \left(\frac{1}{k-1}\right)^{\frac{k-2}{k-1}} \Gamma \left(\frac{1}{k-1},\frac{2 T^{1-k}}{k-1}\right)\Bigg) \\
& \geq &
 \frac{1}{16} t^{k }  
\quad \mbox{for all}\quad t\geq  T_1 \,.
\end{eqnarray*}
The estimate for the first term of (\ref{6.30}) is evident. Lemma is proved. 
\hfill $\square$ 

On the other hand, according to (\ref{3.12a}), we have
\begin{eqnarray*}
F''(t)
& \gtrsim  &
\widetilde{\lambda}  ^{-p} (t)\left( \int_{|x| \le R+ \phi (t)-\phi (1)} |\varphi (x )|^{p/(p-1)} \,dx \right)^{1- p}
t^{1-p}\left| F_1(t) \right|^{p}\quad \mbox{\rm for large }\quad t \,,
\end{eqnarray*}
and, consequently, the asymptotic of $\widetilde{\lambda}(t) $, Lemma~\ref{L3.2},  and Lemma~\ref{L2.3}    imply
\begin{eqnarray*}
F''(t)
& \gtrsim  &
\widetilde{\lambda}  ^{-p} (t)\left( \int_{|x| \le R+ \phi (t)-\phi (1)} |\varphi (x )|^{p/(p-1)} \,dx \right)^{1- p}
t^{1-p}\left| F_1(t) \right|^{p}\\
& \gtrsim  &
c_R     (R+  \phi (t)-\phi (1))^{  -\frac{n-1}{2} (p-2)  } 
t^{1-p} t^{ \frac{pk}{2} }\\
& &
\times \left|    \left\{ \Lambda _1(k)\int_{{\mathbb R}^n}  a(x) u_0 (x)\varphi (x)dx+ \int_{{\mathbb R}^n}  a(x) u_1 (x)\varphi (x)dx\right\}  \right|^{p}\quad \mbox{\rm for }\quad t \geq T\,.
\end{eqnarray*}
Here $T>1$ is a sufficiently large number. It follows
\begin{eqnarray*}
F(t)
& =  &
  F(1)+ \int_1^T   \left\{ F'   ( 1 ) + \int_T^{t_1}F''  (t_2 ) dt_2 \right\} dt_1 + F'(T)(t-T) +
  \int_T^t \int_T^{t_1} F''  (t_2 )
 dt_2 \, dt_1    \\
& \gtrsim  &
  F(1)+  F'   ( 1 ) (T-1) + (t-T)\left\{ F'(1)+\int_1^T F'' (t_1)\, dt_1 \right\}\\
&  &
+
 \left|   \left\{ \Lambda _1(k)\int_{{\mathbb R}^n} a(x)  u_0 (x)\varphi (x)dx+ \int_{{\mathbb R}^n}  a(x) u_1 (x)\varphi (x)dx\right\}  \right|^{p} \\
& &
\times
 \int_T^t \int_T^{t_1} c_R    (R+  \phi (t_2)-\phi (1))^{  -\frac{n-1}{2} (p-2)  }
t_2^{1-p} t_2^{\frac{kp}{2} }  
 dt_2 \, dt_1  \,,
\end{eqnarray*}
where $F'   ( 1 )=  \int_{{\mathbb R}^n}  a(x) u_0 (x)\varphi (x)dx+ \int_{{\mathbb R}^n}  a(x) u_1 (x)\varphi (x)dx$. Thus,
\begin{eqnarray*}
F(t)
& \gtrsim  &
  F(1)+ F'   ( 1 ) (t-1)   
+
 \left| \left\{ \Lambda _1(k) \int_{{\mathbb R}^n}  a(x) u_0 (x)\varphi (x)dx+ \int_{{\mathbb R}^n}  a(x) u_1 (x)\varphi (x)dx\right\}  \right|^{p} \\
& &
\hspace{3.5cm}\times
 \int_T^t \int_T^{t_1}  \phi (t_2) ^{  -\frac{n-1}{2} (p-2)  }
t_2^{1-p}t^{\frac{kp}{2}}_2
 dt_2 \, dt_1   \,.
\end{eqnarray*}
Set
\[
r=  -(1-k)  \frac{n-1}{2} (p-2)   +1-p+\frac{kp}{2}+2\,,\qquad 
q=(p-1)(1+n (1-k))\,.
\]
We need $r\geq  1$,  that is,
\[
  \displaystyle p\leq   2+ \frac{2   k }{n+1-k n }\,.
\]
We check the condition $(p-1)r> q-2$ of the Kato's lemma (see, e.g., \cite[Lemma~2.1]{Yag_2005}), that is, 
\begin{eqnarray*} 
&  &
p^2 (n+1 -k n)- p (2k+n+3-k  n )-2(1-k)    <0\,. 
\end{eqnarray*}  
Since $k<1$, we conclude 
$
1<p<  p_{0}(n,k)$.  
Theorem is proved. \hfill $\square$

\smallskip

For the semilinear generalized Tricomi equation  $   \partial_{t}^2 u-t^m \Delta u=|u|^p$ with   increasing coefficient, that is with $m \in {\mathbb N} $,   the critical exponent $p_{crit}(m,n)$ and conformal exponent $p_{conf}(m,n) $ are suggested in \cite{He-Witt-Yin}. 
Then, there are  interesting articles on the non-linear higher-order degenerate hyperbolic equations \cite{Ruan-Witt-Yin-JDE2014}, the low regularity solution problem for the semilinear mixed type equation \cite{Ruan-Witt-Yin-JDE2015}, and the local existence and singularity structures of
low regularity solution to the semilinear generalized Tricomi equation   with  discontinuous initial data \cite{Ruan-Witt-Yin}.  

The Cauchy problem for the damped linear wave equations 
with  time-dependent propagation speed and dissipations, $u_{tt}-a(t)^2\Delta u + b(t)u_t=0 $, where $a \in L^1(0,\infty) $, is considered in  \cite{Ebert-Reissig}. An interesting example of the quasilinear equation $u_{tt}-t^{-4}\exp( -2t^{-1}) \Delta u - (u_{tt})^2+ t^{-4}\exp( -2t^{-1})(\nabla u)^2=0 $ without global solvability for arbitrarily small initial data is given in \cite{Wirth}. See also \cite{{Choquet-Bruhat-ND_2000},Yag_2005} for more examples of such quasilinear equation.

\section{Local in time solution} 
\setcounter{equation}{0}
\renewcommand{\theequation}{\thesection.\arabic{equation}}
\label{S4}

In this section we prove a local in time existence of the waves propagating  in the Einstein-de~Sitter spacetime.
The initial data are prescribed at the plane $t=0$ where the coefficients are singular. We discuss only the massless fields.
   Denote by $G$ a solution operator of the   problem 
\begin{eqnarray}
\label{ivp_2}
\begin{cases}
 \psi_{tt} - t^{-4/3}A(x,D_x)   \psi +   2   t^{-1}          \psi_t = f,  \qquad t>0 ,\,\, x \in {\mathbb R}^n,\cr
 \displaystyle   \lim_{t\rightarrow 0^+}\, t \psi  (x,t) = \varphi_0 (x),  \quad x \in {\mathbb  R}^n \,,\cr
\displaystyle
\lim_{t\rightarrow 0^+}
  \left(  t \psi_t  (x,t) + \psi  (x,t)+3 t^{ - {1}/{3}}A(x,D_x) \varphi_0   (x   )  \right)
=  \varphi_1 (x),  \quad  x \in {\mathbb  R}^n \,,
\end{cases}
\end{eqnarray}
with $\varphi_0  (x   )=\varphi_1 (x   ) =0$, that is $\psi =G[f] $. Let $\psi _0$ is the solution of the   problem (\ref{ivp_2})  with  $f=0$. 
Then any solution $\psi $ of the problem 
\begin{eqnarray}
\label{ivp_3}
\begin{cases}
 \psi_{tt} - t^{-4/3}A(x,D_x)   \psi +   2   t^{-1}          \psi_t = F(\psi ), 
 \qquad t>0 ,\,\, x \in {\mathbb R}^n,
\cr
 \displaystyle   \lim_{t\rightarrow 0^+}\, t \psi  (x,t) = \varphi_0 (x),  \quad x \in {\mathbb  R}^n \,,\cr
\displaystyle
\lim_{t\rightarrow 0^+}
  \left(  t \psi_t  (x,t) + \psi  (x,t)+3 t^{ - {1}/{3}}A(x,D_x) \varphi_0   (x   )  \right)
=  \varphi_1 (x),  \quad  x \in {\mathbb  R}^n \,,
\end{cases}
\end{eqnarray}
solves also the linear integral equation
\begin{eqnarray}
\label{EdSIE}
\psi (x,t)= \psi _0(x,t) + G[  F (\psi (\cdot ,\tau ))](x,t), \quad t >0\,.
\end{eqnarray}
We define the solution  of (\ref{ivp_3}) as a  solution of the  integral equation (\ref{EdSIE}).
Let $\alpha _0 (n) = (-(n+3) +\sqrt{n^2+30n+81})/(2(n+3))$    be a positive solution of the equation
\[
(n +3)\alpha ^2+(n+3)\alpha -6=0 \,.
\]
\begin{theorem} 
Consider the problem (\ref{ivp_3}) 
for $F(\psi )=|\psi  |^{1+\alpha }  $ or $F(\psi )=|\psi  |^{ \alpha }\psi , $ and with the elliptic operator $A(x,D_x) $ having the properties (\ref{3.12b_intr})-(\ref{3.14_intr}). Assume that  $0<\alpha <\alpha _0 (n)$. For every given $\varphi_0 (x), \varphi_1 (x) $, there exists $T=T(\varphi_0 , \varphi_1 ) $ such that  the problem (\ref{ivp_3}) has a solution $\psi  \in C^2((0,T(\varphi_0 , \varphi_1 )];L^{q} ({\mathbb R}^n))$, where $q=2+ \alpha $.
\end{theorem}
\medskip

\noindent
{\bf Proof.}
  The following estimate  is an analog of \cite{Galstian-Kinoshita-Yagdjian} (see (3.6),(3.7) and Prop.~3.3) and can be proved  by means of 
Theorem~3.1~\cite{Brenner1979}
and the representation formulas of \cite{JDE2015}:
\begin{eqnarray*}
\|
\psi (\cdot ,t)\|_{ L^q  ({\mathbb R}^n)  }
& \le  &
C
 t^{\frac{1}{3}(-1-n(1/p-1/q))} \Big( t^{-\frac{2}{3}}\|\varphi_0  \|_{ L^p ({\mathbb R}^n)  }  +
\|A(x,D_x)   \varphi_0  \|_{ L^p ({\mathbb R}^n)  } \Big)  \\
&  &
+ C t^{\frac{1}{3}(-n(1/p-1/q))}\|\varphi_1    \|_{ L^p ({\mathbb R}^n)  }\\
&  &
+t^{\frac{1}{3}(-n(1/p-1/q))}\int_{ 0}^{t} \tau \| |\psi |^{1+\alpha }(\cdot ,\tau )  \|_{ L^p ({\mathbb R}^n)  }
\,d\tau   \quad \mbox{\rm for all} \quad  t \in (0,T]\,.
\end{eqnarray*}
In particular, for   $q=\alpha +2$ and $p= (\alpha +2)/(\alpha +1) $   we obtain 
\begin{eqnarray*}
t^{1+\frac{n\alpha }{3(\alpha +2)} }
\|\psi (\cdot ,t)\|_{ L^q  ({\mathbb R}^n)  }
& \le  &
C
\left(\|\varphi_0  \|_{ L^p ({\mathbb R}^n)  }  +
\|A(x,D_x)  \varphi_0  \|_{ L^p ({\mathbb R}^n)  } +  \|\varphi_1    \|_{ L^p ({\mathbb R}^n)  } \right)\\ 
&  &
+Ct \int_{ 0}^{t} \tau \| |\psi (\cdot ,\tau ) |^{1+\alpha } \|_{ L^p ({\mathbb R}^n)  }
\,d\tau  \\
& \le  &
C
\left(\|\varphi_0  \|_{ L^p ({\mathbb R}^n)  }  +
\| A(x,D_x)   \varphi_0  \|_{ L^p ({\mathbb R}^n)  } +  \|\varphi_1    \|_{ L^p ({\mathbb R}^n)  } \right)\\ 
&  &
+Ct \int_{ 0}^{t} \tau^{} \| \psi (\cdot ,\tau )  \|_{ L^q ({\mathbb R}^n)  }^{1+\alpha }
\,d\tau   
\end{eqnarray*}  
for all   $t \in (0,T]$. Then, it follows
\begin{eqnarray*}
t^{1+\frac{n\alpha }{3(\alpha +2)} }\|
\psi (\cdot ,t)\|_{ L^q  ({\mathbb R}^n)  }
& \le  &
C
\left(\|\varphi_0  \|_{ L^p ({\mathbb R}^n)  }  +
\|A(x,D_x)  \varphi_0  \|_{ L^p ({\mathbb R}^n)  } +  \|\varphi_1    \|_{ L^p ({\mathbb R}^n)  } \right)\\ 
&  &
+Ct \int_{ 0}^{t} \tau^{-\frac{n\alpha (\alpha +1) }{3(\alpha +2)}-\alpha } \left( \tau ^{1+\frac{n\alpha }{3(\alpha +2)} }\| \psi (\cdot ,\tau )  \|_{ L^q ({\mathbb R}^n)  }\right)^{1+\alpha }
\,d\tau   
\end{eqnarray*}
for all   $t \in (0,T]$. Since $t\psi  $ is continuous at $t=0$ and $\alpha <\alpha (n) $, we obtain
\begin{eqnarray*}
t^{1+\frac{n\alpha }{3(\alpha +2)} }\|
\psi (\cdot ,t)\|_{ L^q  ({\mathbb R}^n)  }
& \le  &
C
\left(\|\varphi_0  \|_{ L^p ({\mathbb R}^n)  }  +
\|A(x,D_x)   \varphi_0  \|_{ L^p ({\mathbb R}^n)  } +  \|\varphi_1    \|_{ L^p ({\mathbb R}^n)  } \right)\\ 
&  &
+Ct \max_{\tau \in[0,t]} \left( \tau ^{1+\frac{n\alpha }{3(\alpha +2)} }\| \psi (\cdot ,\tau )  \|_{ L^q ({\mathbb R}^n)  }\right)^{1+\alpha }\int_{ 0}^{t} \tau^{-\frac{n\alpha (\alpha +1)}{3(\alpha +2)}-\alpha } 
\,d\tau  
\end{eqnarray*}
for all   $t \in (0,T]$. Hence, for $\alpha <\alpha _0 (n) $ we have
\begin{eqnarray*}
t^{1+\frac{n\alpha }{3(\alpha +2)} }\|
\psi (\cdot ,t)\|_{ L^q  ({\mathbb R}^n)  }
& \le  &
C
\left(\|\varphi_0  \|_{ L^p ({\mathbb R}^n)  }  +
\|A(x,D_x)  \varphi_0  \|_{ L^p ({\mathbb R}^n)  } +  \|\varphi_1    \|_{ L^p ({\mathbb R}^n)  } \right)\\ 
&  &
+Ct^{2-\frac{n\alpha (\alpha +1) }{3(\alpha +2)}-\alpha }  \max_{\tau \in[0,t]} \left( \tau ^{1+\frac{n\alpha }{3(\alpha +2)} }\| \psi (\cdot ,\tau )  \|_{ L^q ({\mathbb R}^n)  }\right)^{1+\alpha }
\,d\tau    
\end{eqnarray*}
for all   $t \in (0,T]$. If we consider the map $S  $ defined as follows
\[
S[ \psi](x,t) := \psi _0(x,t)+  G[   |\psi (\cdot ,\tau )|^p](x,t), \quad \forall t \in [0,T]\,,
\]
then the last estimate   implies that $S$ is a contraction for small $T$. Indeed, for $\psi_1 $ and $\psi_2 $ we obtain 
\begin{eqnarray*}
&  &
\max_{t \in [0, T]}  t^{1+\frac{n\alpha }{3(\alpha +2)} }\|
\psi_1 (\cdot ,t)- \psi_2 (\cdot ,t)\|_{ L^q  ({\mathbb R}^n)  }   \\
& \le  & 
c \max_{t \in [0, T]}\left(  t^{1+\frac{n\alpha }{3(\alpha +2)} }\|
\psi_1 (\cdot ,t)- \psi_2 (\cdot ,t)\|_{ L^q  ({\mathbb R}^n)  } \right)^{\alpha+1}
T^{2-\frac{n\alpha (\alpha +1)}{3(\alpha +2)}-\alpha }\,.
\end{eqnarray*}
The theorem is proved. \hfill $\square$

Thus, for $n=3$ we have the following range of $\alpha  $ of the nonlinear term $\alpha +1  \in(1,(\sqrt{5}+1)/2 )$. 
\bigskip

\section{Uniqueness. Finite speed of propagation property}
\label{S5}
\setcounter{equation}{0}

In \cite{Galstian-Kinoshita-Yagdjian} and \cite{JDE2015} the   representations  for the solutions  of the initial value problem for the equations with   singular coefficients are given. 
Because of that particular type of  singularity in the coefficients one cannot apply the known uniqueness theorems (see, e.g., \cite{B-G}).
The uniqueness  must  be established independently of the   representation formulas.  For the case of $A(x,D_x)= \Delta  $ it was done in \cite{Galstian-Yagdjian-MPAG}. 
In this section  we prove the  uniqueness of the solution and then the finite speed of propagation property.

Suppose that
\[
A(x,D_x)= \sum _{|\alpha | \leq 2} a_\alpha (x)\partial_x^\alpha 
\]
is negative elliptic operator with smooth coefficients $ a_\alpha (x) \in C^\infty({\mathbb R}^n)$ such that 
\[
A(x,D_x)= A(\infty,D_x) \quad \mbox{\rm for all}  \quad x \in {\mathbb R}^n \quad |x|\geq R_A\,,
\]
and
\[
\sum _{|\alpha | = 2} a_\alpha (x)\xi ^\alpha > 0 \quad if \quad \xi \in {\mathbb R}^n, \quad \xi \not= 0\,,\quad x \in {\mathbb R}^n.
\]

\begin{theorem}
\label{Tuniq}
Assume that $ A(x,D_x)$ is elliptic negative self-adjoint operator. The solution $\psi $ of the problem 
\begin{eqnarray}
\label{ivp}
\begin{cases}
 \psi_{tt} - t^{-4/3}A(x,D_x)    \psi +   2   t^{-1}          \psi_t = f,  \qquad t>0 ,\,\, x \in {\mathbb R}^n,\cr
 \displaystyle   \lim_{t\rightarrow 0^+}\, t \psi  (x,t) = \varphi_0 (x),  \quad x \in {\mathbb  R}^n \,,\cr
\displaystyle
\lim_{t\rightarrow 0^+}
  \left(  t \psi_t  (x,t) + \psi  (x,t)+3 t^{ - {1}/{3}} A(x,D_x) \varphi_0   (x   )  \right)
=  \varphi_1 (x),  \quad  x \in {\mathbb  R}^n \,,
\end{cases}
\end{eqnarray} 
is unique in $   C^2((0,T]; {{\mathcal{D}}}' ({\mathbb R}^n)) $.
\end{theorem}
\medskip

\noindent
{\bf Proof.} It suffices to prove the uniqueness in the problem 
\begin{eqnarray*} 
\begin{cases}
\vspace{0.2cm}   u_{tt} - t^{-4/3} A(x,D_x)   u  = 0,  \qquad  \quad \mbox{in}\quad  t>0 ,\,\, x \in {\mathbb R}^n,\cr 
 \displaystyle    u (x,0) =   0  , \qquad  
  u _t (x,0) 
=  0,   \quad \mbox{in}\quad  {\mathbb R}^n  \,,
\end{cases}
\end{eqnarray*} 
where $ u=t \psi  \in C^1([0,T]; {{\mathcal{D}}}' ({\mathbb R}^n))\cap  C^2((0,T]; {{\mathcal{D}}}' ({\mathbb R}^n)) $.  
We choose an arbitrary $T>0$ and for the function $\varphi  \in C_0^\infty( {\mathbb R}^n) $ consider the Cauchy problem
\begin{eqnarray}
\label{CPT} 
\begin{cases}
\vspace{0.2cm}   v_{tt} - t^{-4/3} A(x,D_x)  v = 0,  \qquad  \quad \mbox{in}\quad  t\in(0,T] ,\,\, x \in {\mathbb R}^n,\cr 
 \displaystyle    v (x,T) =   0  , \qquad  
  v _t (x,T) 
= \varphi (x),   \quad \mbox{in}\quad  {\mathbb R}^n   . 
\end{cases}
\end{eqnarray}
Since the operator ${\mathcal S}=\partial_{t}^2 - t^{-4/3} A(x,D_x)$ is strictly hyperbolic for $t>0$, there is a 
unique solution $v \in C^\infty((0,T]\times {\mathbb R}^n))$. 
This solution obeys  finite speed of propagation, consequently 
there is a ball $ B \subseteq {\mathbb R}^n$ of the finite radius $R $, such that 
supp$\,v \subseteq  [0,T] \times B$.

Then we define operator $ \sqrt{-A(x,D_x)}$ (see, e.g.,\cite[Ch.XII]{Taylor}), which is a pseudodifferential operator. The solution $v(x,t) $ of the problem (\ref{CPT}) can be written in terms of the Fourier integral operators as follows
 \begin{eqnarray*}   
v(x,t)  
& = &
  \frac{i}{18} (\sqrt{-A(x,D_x)}) ^{-3}\\
&  &
\times \Bigg\{ 
\left(i   \phi (t) \sqrt{-A(x,D_x)}-1 \right) \left( i  \phi (T) \sqrt{-A(x,D_x)}+1 \right) e^{-i    \left(\phi (T) - \phi (t)\right) \sqrt{-A(x,D_x)}}\\
&  &
-\left(i  \phi (t) \sqrt{-A(x,D_x)}+ 1 \right) \left(i  \phi (T) \sqrt{-A(x,D_x)}-1 \right) e^{i   \left(\phi (T) - \phi (t)\right) \sqrt{-A(x,D_x)}}  \Bigg\}\varphi (x)
\end{eqnarray*} 
as well as 
 \begin{eqnarray*}  
 &  &
v(x,t) \\
& = &
  \frac{1}{9} (\sqrt{-A(x,D_x)}) ^{-3}\Bigg\{ \left( \phi (t) \phi (T) \sqrt{-A(x,D_x)}  -1\right) \sin \left(\sqrt{-A(x,D_x)} \left(\phi (T)- \phi (t)\right)\right)\\
&  &
- \sqrt{-A(x,D_x)} \left(\phi (t)- \phi (T)\right)  \cos \left(\sqrt{-A(x,D_x)} \left(\phi (T)- \phi (t)\right)\right) \Bigg\}\varphi (x)\,.
\end{eqnarray*} 
Thus, the solution is given by the Fourier integral operators of order $-1$. 
In particular, for the derivative we obtain
 \begin{eqnarray*}   
v_t(x,t) 
& = &
  \frac{1}{2\phi (t)  }  (\sqrt{-A(x,D_x)})^{-1}
\Bigg\{ e^{ -i \left( \phi (t) -\phi (T)\right)\sqrt{-A(x,D_x)}}  \left(\sqrt{-A(x,D_x)} \phi (T)+i\right)  \\
&  &
\hspace{4cm} +e^{i \left( \phi (t) -\phi (T)\right)\sqrt{-A(x,D_x)}}\left( \sqrt{-A(x,D_x)} \phi (T)-i\right) \Bigg\} \varphi (x)\\
& = &
  \frac{1}{\phi (t)  } \Bigg\{ \cos \left( \left( \phi (t) -\phi (T)\right)\sqrt{-A(x,D_x)}  \right)  \phi (T)\\
&  &
\hspace{3cm} +   (\sqrt{-A(x,D_x)})^{-1}\sin \left(  \left( \phi (t) -\phi (T)\right)\sqrt{-A(x,D_x)} \right)    \Bigg\} \varphi (x)\,.
\end{eqnarray*}
One can easily check the following limits
 \begin{eqnarray*}   
\lim _{t \to 0^+} v(x,t)  
& = &
 \frac{1}{18}(\sqrt{-A(x,D_x)}) ^{-3} \\
&  &
\times \Bigg\{ e^{-  i \sqrt{-A(x,D_x)} \phi (T)}   \sqrt{-A(x,D_x)} \phi (T)-ie^{-  i \sqrt{-A(x,D_x)} \phi (T)} \\
&  &
+e^{  i \sqrt{-A(x,D_x)}  \phi (T)}  \sqrt{-A(x,D_x)}  \phi (T)+ie^{  i \sqrt{-A(x,D_x)}  \phi (T)}  \Bigg\}\varphi  (x)\\ 
& = &
 -\frac{1}{9  } (A(x,D_x))^{-1} \Bigg\{\cos \left(  \phi (T) \sqrt{-A(x,D_x)}\right)   \phi (T) \\
&  &
\hspace{3cm}-  (\sqrt{-A(x,D_x)}) ^{-1} \sin \left(   \phi (T)   \sqrt{-A(x,D_x)}\right)  \Bigg\}\varphi  (x)\\
\end{eqnarray*} 
and
\begin{eqnarray*}
&  &
\lim _{t \to 0^+} \phi (t)v_t(x,t)  \\
 & =  &
\left\{\phi (T)  \cos \left( \phi (T)\sqrt{-A(x,D_x)}  \right)+(\sqrt{-A(x,D_x)}) ^{-1}  \sin \left( \phi (T)\sqrt{-A(x,D_x)} \right)\right\}\varphi  (x)\,.
\end{eqnarray*}  
In particular, it follows 
\[
v, \, t^{1/3}v_t \in C([0,T]; C^\infty(K))\,,
\]
where $K\subseteq  {\mathbb R}^n $ is a compact. We denote $\langle u,\varphi \rangle $ the pairing of the distribution $u \in {\mathcal D}'({\mathbb R}^n) $ and a test function $\varphi  \in {\mathcal D}({\mathbb R}^n) $. Consider the functions $\langle  u    ( \cdot , t ),v ( \cdot , t ) \rangle  $, 
$\langle u_{ t}   ( \cdot , t ),v_t ( \cdot , t ) \rangle $, and 
$\langle  u  ( \cdot , t ),A(\cdot ,D_x) v ( \cdot , t )\rangle  $. We can assume that  supp\,$u \subseteq [0,T]\times B_u $, where $B_u$ is a compact and  it contains   \,$B$. Then  we can   estimate  these functions as follows
\[
|\langle  u    (\cdot , t ),v ( \cdot , t ) \rangle  |  +  |\langle  u    (\cdot , t ),A(\cdot ,D_x) v ( \cdot , t ) \rangle  |\leq c t \quad \mbox{\rm for all} \quad t \in [0,T]\,.
\]
Hence,
\[
\int_0^T t^{-4/3}  |< u    (\cdot , t ),A(\cdot ,D_x) v ( \cdot , t ) > | \,dt < \infty \quad \mbox{\rm and} 
\quad  \int_0^T  |  < u_{tt}     ( \cdot , t ),v ( \cdot , t ) >|\,dt  < \infty\,,
\]
as well as
\[
 \int_0^T |< u_{t}     ( \cdot , t ),v_t ( \cdot , t ) > |\, dt < \infty  \,.
\]
Hence, taking into account that  $u $  solves equation without source term, we obtain
\[
\int_0^T    < u_{tt}     ( \cdot , t ),v ( \cdot , t ) >\,dt
- \int_0^T t^{-4/3}  < u     ( \cdot , t ), A(\cdot ,D_x)v ( \cdot , t ) >\,dt =0 \,.
\]
Applying the  integration by parts, taking into account that $v$ solves equation without source term, we derive 
\[
< u    (\cdot ,T ),\varphi (\cdot )>  =0 
\]
for arbitrary $\varphi  \in C_0^\infty( {\mathbb R}^n) $, which completes the proof of the theorem. 
\hfill $\square$
\bigskip

Theorem~\ref{Tuniq} allows us to prove the finite speed of propagation property in the Cauchy problem.
\begin{theorem}
The solution $\psi  \in   C^2((0,T]; {\mathcal{D}}' ({\mathbb R}^n)) $
of the problem (\ref{ivp}) obeys  finite speed of propagation, that is, for every given $T>0$ and the open ball $B_R(x_0)=\{x \in {\mathbb R}^n\,;\, |x-x_0| <R\} $, if 
\[
 \varphi_0 =\varphi_1=0  \quad \mbox{on}\quad   B_{R+ 3T^{1/3}s_A}(x_0)  \quad \mbox{and}\quad    f=0   \quad \mbox{on}\quad   \bigcup _{t \in [0,T]} B_{R+ 3(T^{1/3}-t^{1/3})s_A} (x_0) \, , 
\]
then 
\[
\psi (T ) =0  \quad \mbox{on}\quad  B_R(x_0)=\{ x \in {\mathbb R}^n\,;\, |x-x_0| <R\}  \,.
 \]
 Here
 \[
 s_A = \max_{ x \in {\mathbb R}^n, \, \xi  \in {\mathbb R}^n ,\,  |\xi|=1}   \sum _{|\alpha | = 2} a_\alpha (x)\xi ^\alpha  \,.
 \]
 \end{theorem}
 \medskip
 
 \noindent
 {\bf Proof.} It suffices  to use  the  finite speed of propagation in the problem for the auxiliary function $v$ in the proof of the previous theorem.
 \hfill $ \square$

\section*{Acknowledgement}
This paper was completed during our visit at the Technical University Bergakademie
Freiberg in the summer of 2016. The authors are grateful to Michael Reissig for the invitation
to Freiberg and for the warm hospitality.  K.Y. expresses his gratitude to the Deutsche
Forschungsgemeinschaft for the financial support under the grant GZ: RE 961/21-1. The  authors are grateful to Alessandro Palmieri for the useful remark that improved the text of the manuscript.

\end{document}